\documentclass[sigconf]{acmart}

\AtBeginDocument{%
  }

\setcopyright{acmlicensed}
\copyrightyear{2018}
\acmYear{2018}
\acmDOI{XXXXXXX.XXXXXXX}

\acmConference[Conference acronym 'XX]{Make sure to enter the correct
  conference title from your rights confirmation emai}{June 03--05,
  2018}{Woodstock, NY}
\acmISBN{978-1-4503-XXXX-X/18/06}

\usepackage{graphicx}
\usepackage{amsmath}
\usepackage{multirow}
\usepackage{booktabs}
\usepackage{array}
\usepackage{subfigure}
\usepackage{amssymb}
\usepackage{makecell}
\usepackage{verbatim}
\usepackage{bm}     % for \bm
\usepackage{enumitem}       % for itemize
\usepackage{xcolor}
\usepackage{algpseudocode}

\begin{document}

\title{DFGNN: Dual-frequency Graph Neural Network for Sign-aware Feedback}

\author{Yiqing Wu}
% \authornote{Yiqing Wu is also at the University of Chinese Academy of Sciences, China.}

\affiliation{%
  \institution{Institute of Computing Technology, Chinese Academy of Sciences}
  \city{Beijing}
  \country{China}}
\email{iwu_yiqing@163.com}

\author{Ruobing Xie}
\affiliation{%
  \institution{Tencent}
  \city{Beijing}
  \country{China}
}
\email{ruobingxie@tencent.com}

\author{Zhao Zhang}

\affiliation{%
  \institution{Institute of Computing Technology, Chinese Academy of Sciences}
  \city{Beijing}
  \country{China}
}

\email{zhangzhao2021@ict.ac.cn}
\author{Xu Zhang}
\affiliation{%
  \institution{Tencent}
  \city{Beijing}
  \country{China}
}
\author{Fuzhen Zhuang}
\affiliation{%
  \institution{Institute of Artificial Intelligence, Beihang University}
  \city{Beijing}
  \country{China}
}
% \authornote{Fuzhen Zhuang is also at  Zhongguancun Laboratory, Beijing, China
% }
\email{zhuangfuzhen@buaa.edu.cn}

\author{Leyu Lin}
\author{Zhanhui Kang}
\affiliation{%
  \institution{Tencent}
  \city{Beijing}
  \country{China}
}
\email{{goshawklin, kegokang}@tencent.com}
\author{Yongjun Xu}
\affiliation{%
  \institution{Institute of Computing Technology, Chinese Academy of Sciences}
  \city{Beijing}
  \country{China}
}
\email{xyj@ict.ac.cn}
\renewcommand{\shortauthors}{Trovato et al.}

\begin{abstract}
The graph-based recommendation has achieved great success in recent years. However, most existing graph-based recommendations focus on capturing user preference based on positive edges/feedback,  while ignoring negative edges/feedback (e.g., dislike, low rating) that widely exist in real-world recommender systems. How to utilize negative feedback in graph-based recommendations still remains underexplored. In this study, we first conducted a comprehensive experimental analysis and found that (1) existing graph neural networks are not well-suited for modeling negative feedback, which acts as a high-frequency signal in a user-item graph. (2) The graph-based recommendation suffers from the representation degeneration problem. Based on the two observations, we propose a novel model that models positive and negative feedback from a frequency filter perspective called  Dual-frequency Graph Neural Network for Sign-aware Recommendation (DFGNN). Specifically, in DFGNN, the designed dual-frequency graph filter (DGF)  captures both low-frequency and high-frequency signals that contain positive and negative feedback. Furthermore, the proposed signed graph regularization is applied to maintain the user/item embedding uniform in the embedding space to alleviate the representation degeneration problem. Additionally, we conduct extensive experiments on real-world datasets and demonstrate the effectiveness of the proposed model. Codes of our model will be released upon acceptance.

\end{abstract}




\received{20 February 2007}
\received[revised]{12 March 2009}
\received[accepted]{5 June 2009}

\maketitle
\section{Introduction}

Personalized recommender systems aim to provide appropriate items to the users on platforms. In such systems, the interactions between users and items can naturally be regarded as a user-item bipartite graph. Inspired by the success of Graph Neural Networks (GNNs), the graph-based recommendation model has been widely studied and achieved remarkable progress. However, most existing graph-based recommendation models are merely built on positive edges/feedback. Actually, there is always various negative feedback in real-world recommender systems, such as a low rating, dislike, and short watching time. The negative feedback generally represents what people do not like or even hate.  Recommendations of such items to users may significantly harm the user experience. Besides, only focusing on what users like while ignoring what users dislike may lead to homogenization in the recommended items \cite {zhang2018coupledcf}.

Although negative feedback is crucial and is believed to benefit the recommendation \cite{jeunen2019revisiting}, how to model the negative feedback in graph-based recommendation still remains underexplored. A trivial solution is to introduce negative feedback edges to the graph and model them in a manner similar to positive feedback. However, existing GNNs heavily rely on the homophily assumption, where two nodes connected by an edge tend to be more similar. While, for negative feedback, two nodes connected by an edge may imply dissimilarity. 
%Therefore, employing traditional GNNs to model negative feedback may be inappropriate.
A similar field is the signed graph neural network. It is designed for processing signed graphs, which contain two opposite edge types (e.g., accept/reject, fraud/trustworthiness). The signal in the signed graph generally is objective, and most of them are built upon balance theory, which means the friend of my friend is my friend, and the enemy of my enemy is my friend.
However, the user feedback in recommendations is subjective and personalized, and the balance theory does not always hold in recommendations \cite{seo2022siren}.
Currently, limited works \cite{huang2023negative} have studied the negative feedback in graph-based recommendations. While they still adopt GNNs that rely on the homophily assumption.

Based on the above analysis, we start to ponder \emph{"How to model positive and negative feedback in graph-based recommendation?"} To this end, we deeply contemplate the characteristics of positive and negative feedback.
(1)\textit{ Positive and negative feedback carry different meanings.}
Generally, if an item fits users' interests, the people will produce positive feedback.  In recommendation, this implies a certain degree of similarity between items and users. In contrast, negative feedback implies dissimilarity.  While this makes the negative feedback violate the homophily assumption of GNN. 
(2) \textit{The reasons for negative feedback are heterogeneous.}
Compared to positive feedback, the reasons for users produce negative feedback are diverse and complex.  It could stem from a variety of reasons such as the user hating the categories of the item, the items not meeting the user's expectations, or even due to the user hating a specific word in the title.  
(3)\textit{The negative feedback is generally sparse.} The recommender system aims to recommend what people will like, the negative feedback actually is an anomaly, making it sparse. For example, in the Amazon dataset, less than 10\% of reviews have low ratings.

Looking deeper into the structure information of graph-based recommendation, we can discover that: a user node is connected to many nodes that are similar to it by positive feedback, resulting in a smooth similarity terrain. However, there are also a few strange peaks that stand out, appearing out of place, that is the negative feedback. This inspires us to think about positive and negative feedback essentially from the perspective of signal processing. In which the smooth and unchanged signals are regarded as low-frequency signals. While steep, rapid changes are considered as high-frequency signals. This is highly analogous to the characteristics of positive and negative feedback. To evaluate our assumption, we utilize 
graph spectral theory to 
experimentally analyze the positive and negative feedback in the frequency-domain (detailed in Sec.~\ref{sec:frequency-domain anlayze}) and find that the positive feedback indeed implies low-frequency signals in the graph while the negative feedback implies high-frequency signals in the graph.
Furthermore, the GNNs tend to over-smooth low-frequency signals. Through experiment, we also find that the learned embedding suffers from a representation degeneration problem, which means the embedding shrinks to a small piece of embedding space thus resulting in a loss of expressive power. This issue has a more significant impact on signed graphs, as it leads to an inability to distinguish between negative and positive feedback.

Considering the characteristics of negative and positive feedback, we propose a novel method to model them in graph-based recommendations, called \textbf{Frequency-aware signed graph neural network (DFGNN)}. Specifically, considering the low-frequency signal and high-frequency signal contained in positive and negative feedback, respectively, we propose a dual-frequency graph filter(DGF). In DGF we utilize the designed low-passing graph filter to model the positive feedback while adopt the high-passing filter to model the negative feedback.  To alleviate the representation degeneration problem caused by GNN, we propose a signed graph regularization (SGR) loss. SGR  considers both the similar and dissimilar information in the signed graph and regularizes the learned embedding to be uniformly distributed in the space.
The experiments on real-world datasets demonstrate that our model achieves significant improvements over competitive baselines.

Overall, the contribution of this paper is summarized as follows:
\begin{itemize}
    \item We first conduct a comprehensive analysis of positive and negative feedback in the graph-based recommendation from the frequency perspective. Furthermore, we design a dual-frequency filter to both model low-frequency signals and high-frequency signals contained in positive and negative feedback.
    \item We investigate the representation degeneration problem in graph recommendation, and propose a sign graph regularization loss to alleviate it.
    \item We conduct extensive experiments on real-world datasets and evaluate the effectiveness of the proposed DFGNN in two recommendation tasks. 
\end{itemize}

% 再现实世界中不仅有着正反馈还有负反馈， 负反馈很重要
% 尽管负反馈的建模十分重要但是如何在图推荐中利用负反馈信号仍然缺乏探索。一个相似的领域是signed graph neural network, 在singed graph 中有着两种相反信号的边，例如支持/反对 真/假 敌人/朋友 在推荐中不合适。 
%其次图神经网络往往会遭受到过平滑的问题。

\section{Preliminaries}
\subsection{Signed Graph in Recommendation} In real-world online recommender systems, users generally have various interactions with items and generate feedback to the system, including both positive feedback (e.g. like, buy) and negative feedback (e.g. dislike, low rating). Based on the positive and negative feedback, we define the signed graph in recommendation.  Specifically, we denote the signed graph as $G=\{U, V, \{\mathcal{E}^{+},\mathcal{E}^{-} \}\}$, where $U=\{u_1,u_2, ... , u_{|U|}\}$ and $V=\{v_1,v_2, ... , v_{|V|}$ are the user set and item set in the system, and $ \{\mathcal{E}^{+},\mathcal{E}^{-}\}$ are positive and negative edges in the graph, respectively. If a user has positive feedback for an item, there will be a positive edge $\xi=(u,v,+) \in \mathcal{E}^{+} $. Conversely, a negative edge $\xi=(u,v,+) \in \mathcal{E}^{-} $ will be created if there is negative feedback.

% \subsection{Object} Given a signed graph in the recommendation, our goal is to learn a  function that can map each user and item node to representation $e^{u}_i$  and  $e^{v}_i$ that can be used to downstream recommendation task. 

% \subsection{Graph Convolution  from Filter Perspective }
% In this section, we analyze the graph convolution network from a filter perspective to better understand the role that graph convolution plays in graph learning. In the following we first introduce the graph Fourier transform, then we prove that classical GCN is a low-passing filter. 
\subsection{Graph Fourier Transform}
To better understand the graph Fourier transform. We first introduce the Fourier transform.
Fourier transform is a powerful analytical tool and is widely applied across various fields. It transforms signals from the time domain to the frequency domain. The Fourier transform can be formulated as :
\begin{equation}
f(\omega) = \mathcal{F}(f(x),\omega) \int_{-\infty}^{\infty} f(x) \cdot e^{-2\pi i \omega} \, dx,
\end{equation}
\begin{figure}
\setlength{\abovecaptionskip}{0cm} 
        \setlength{\belowcaptionskip}{0pt}
\captionsetup[subfigure]{skip=0pt,margin=0pt}
  \centering
\captionsetup[subfigure]{skip=0pt,margin=0pt}
  \subfigure[sin(x)]{
    \includegraphics[width=0.31\linewidth]{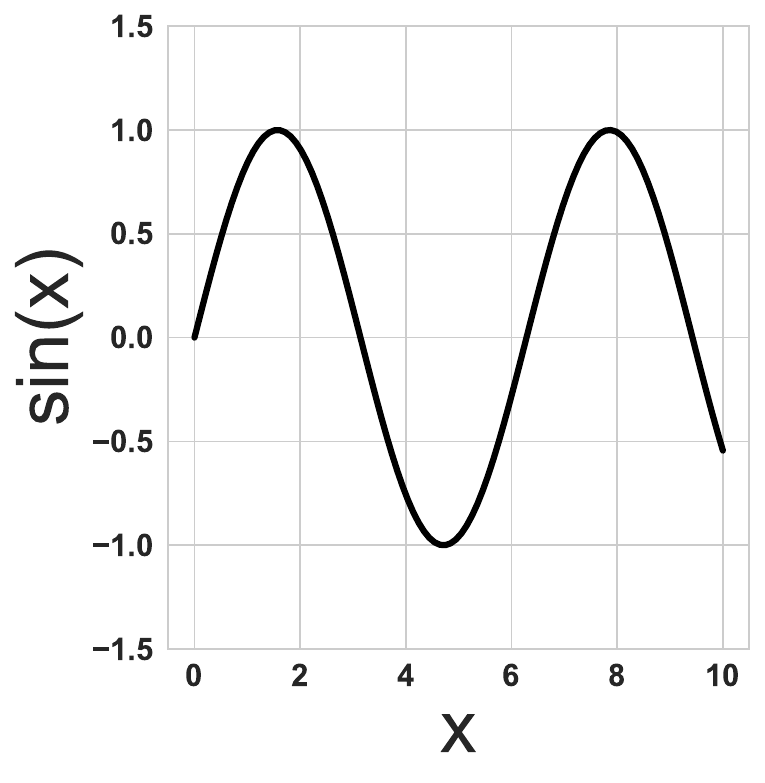}
  }
  \hfill
  \subfigure[0.2\,sin(10x)]{
    \includegraphics[width=0.30\linewidth]{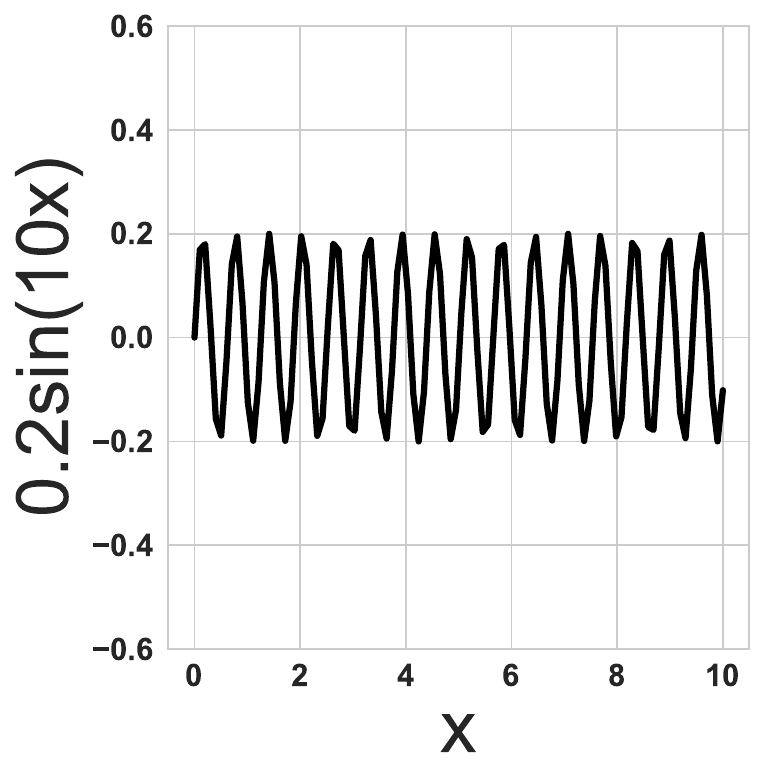}
  }
 \hfill
  \subfigure[0.2\,sin(10x)+sin(x)]{
    \includegraphics[width=0.31\linewidth]{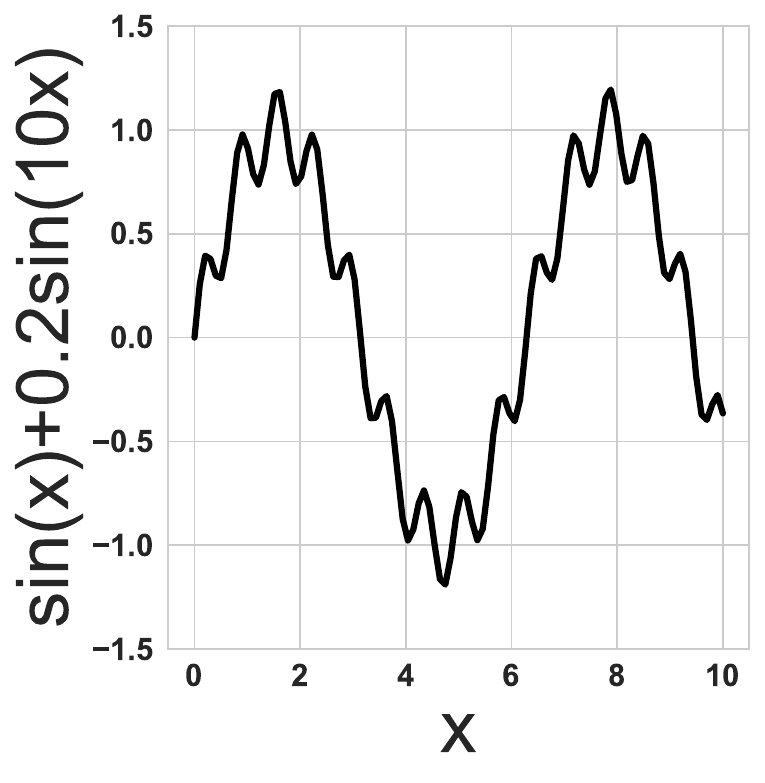}
  }
   \vfill
    \subfigure[$\mathcal{F}$\,(sin(x))]{
    \includegraphics[width=0.31\linewidth]{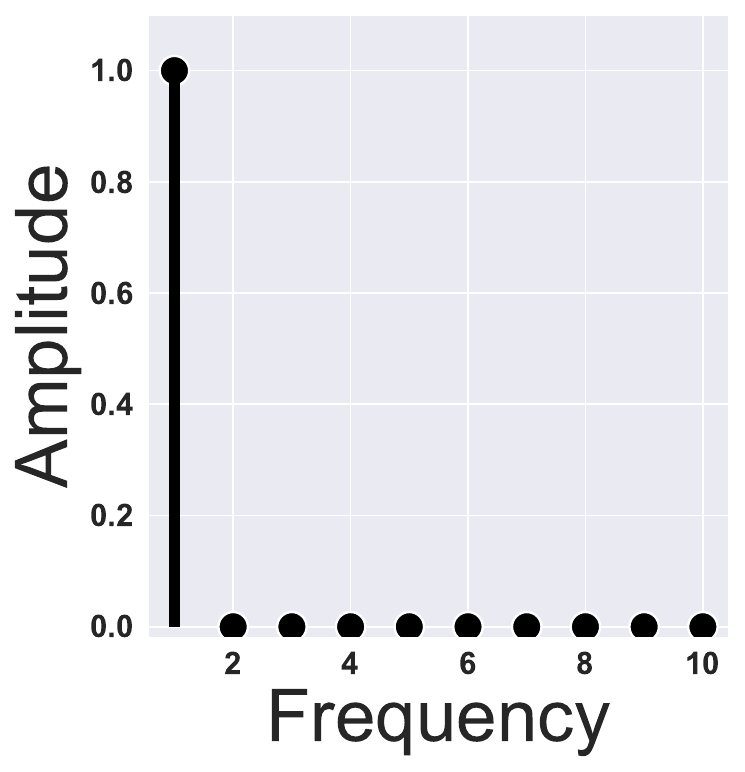}
  }
  \hfill
  \subfigure[$\mathcal{F}$\,(0.2\,sin(10x))]{
    \includegraphics[width=0.30\linewidth]{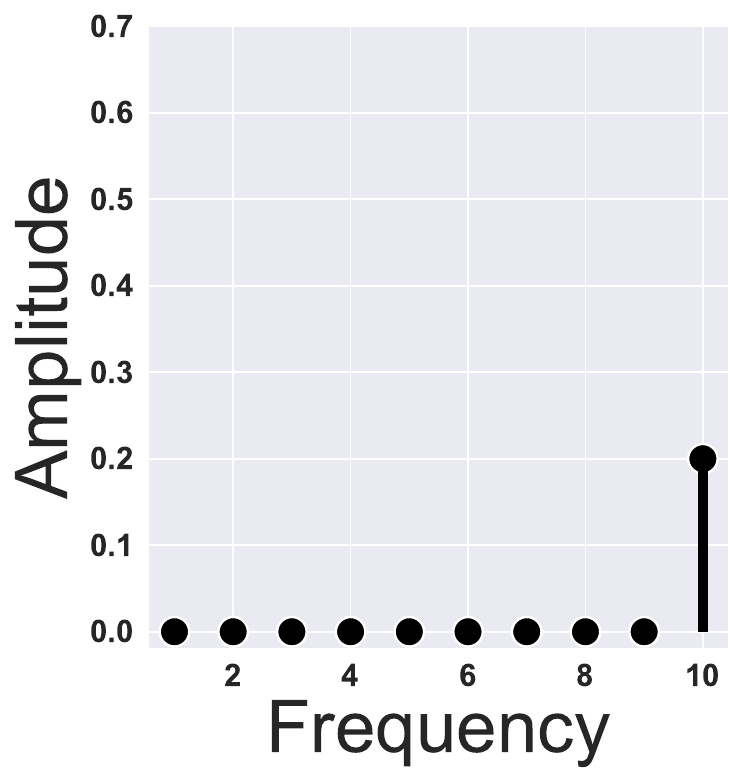}
  }
 \hfill
  \subfigure[$\mathcal{F}$\,(0.2\,sin(10x)+sin(x))]{
    \includegraphics[width=0.31\linewidth]{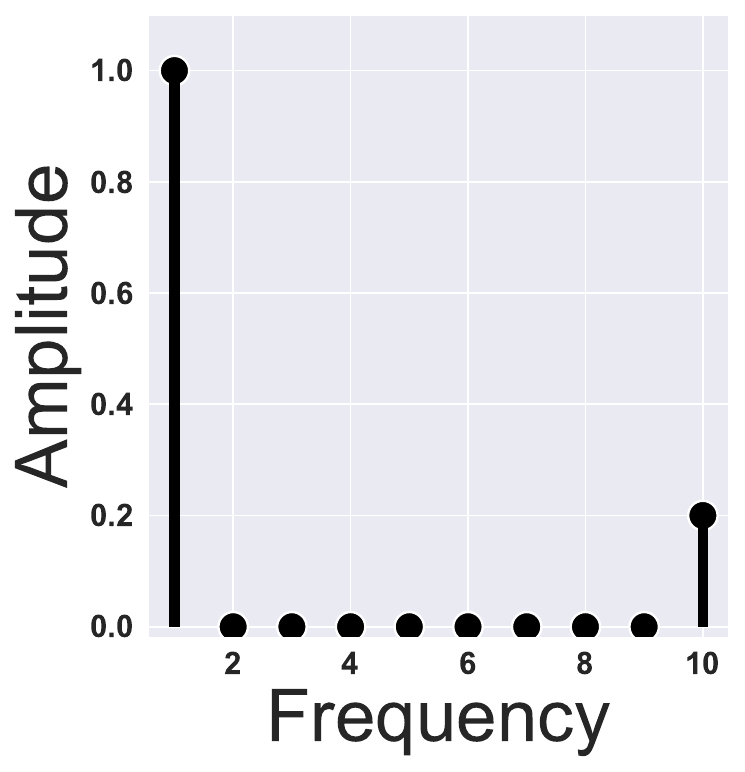}
  }
   
\caption{Different frequency signals in the time domain and frequency domain.}
\label{fig:Frequency}
\end{figure}
where $\mathcal{F}$ is denoted as Fourier transform operation,  $\omega$ is the frequency, $f(\omega)$ are the amplitudes  of corresponding $\omega$.  \textbf{Generally, low frequencies correspond to smooth signals, while high frequencies correspond to non-smooth signals. The amplitude represents the strength of the signal at the corresponding frequency.} To better understand this, we demonstrate this phenomenon in Fig.\ref{fig:Frequency}. We can see that $sin(x)$ (i.e., Fig.~\ref{fig:Frequency}(a)) is smoother, thus its corresponding low frequency is higher(i.e., Fig.~\ref{fig:Frequency}(d)). In contrast, $0.2sin(10x)$ oscillates rapidly (i.e., Fig.~\ref{fig:Frequency}(b)), and correspondingly, the amplitude of high frequencies is higher (i.e. Fig.~\ref{fig:Frequency}(e)).  Fig.~\ref{fig:Frequency}(c)  looks complex in the time domain, but in the frequency domain (Fig.~\ref{fig:Frequency}(f)), it looks simple—it's just the superposition of two sine waves, $sin(x)$ and $0.2sin(10x)$.

To extend the Fourier transform to the graph/non-euclidean domain, Laplacian matrices ${L}$ are introduced to the graph processing field. Classical Laplacian matrice is defined as $L=D-A$, where $A$ is the adjacent matrix and $D$ is the degree matrix of a graph. $L$ is a real symmetric matrix, thus it can be written as $L=E\Lambda E^T$, where $E=\{e_1,e_2, ..., e_N\}$ is the eigenvector and  $\Lambda=\{\lambda_1,\lambda_2, ..., \lambda_N\}$ is the eigenvalue of $L$. Similar to the Fourier transform in the time domain, the Fourier transform in the graph/non-euclidean domain is defined as:
\begin{equation}
f(\lambda_l) = \sum_{0}^N x(i)e^{\lambda_l}(i),
\end{equation}
where $e^{\lambda_l}$ is the eigenvector that corresponds to eigenvalue $\lambda_l$, and $x(i)$ is the node feature of $i$-th node. Graph Fourier Transform shares similar characteristics with the time-domain Fourier Transform. \textbf{Low $\lambda$ values represent low-frequency signals, while high $\lambda$ values represent high-frequency signals}. Besides, $f(\lambda)$ represents the strength of the corresponding frequency.

% 图傅里叶变换可以定义为xxx。其中 w 代表频率， ，一般低的w代表平平滑的信号，高的w代表 不平滑的信号 f(w) 代表振幅，如果对应的振幅高 那么xxx 如果对应的振幅低那么xxx
% 图的卷积定义, 介绍图拉普拉斯

\section{Experimental Observations}
In this section, we aim to analyze the characteristics of negative and positive feedback and the challenges during the processing of signed graphs in recommendation. In the following, we first analyze the negative feedback from a frequency perspective, then we analyze the representation degeneration problem in the graph-based recommendation.

\subsection{Analysis of Positive and Negative Feedback from  Frequency-domain}
\label{sec:frequency-domain anlayze}
\begin{figure}
  \centering
\setlength{\abovecaptionskip}{0cm} 
        \setlength{\belowcaptionskip}{0pt}
  \subfigure[negative feedback]{
    \includegraphics[width=0.46\linewidth]{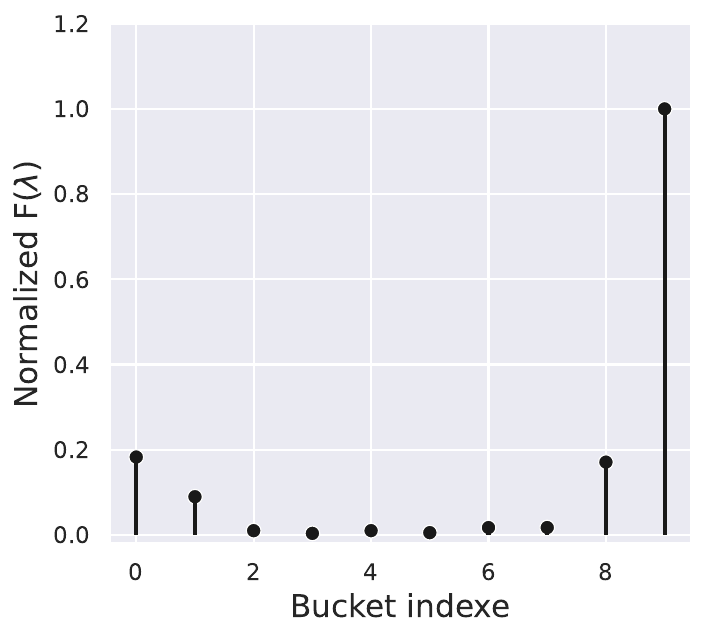}
  }
  \hfill
  \subfigure[positive feedback]{
    \includegraphics[width=0.46\linewidth]{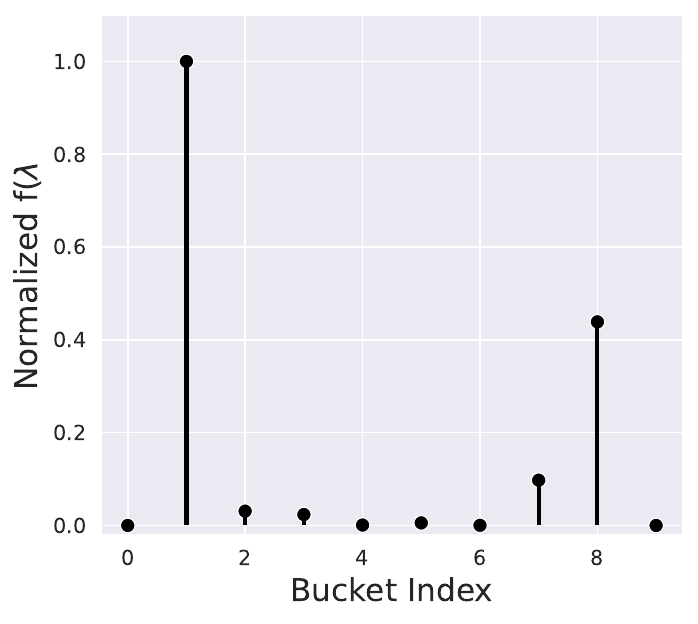}
  }

  \caption{The distribution of f($\lambda$) calulated based on normalzied Laplacian matrix. (a) It is the distribution of only negative edges graph. (b) It is the distribution of only positive edges graph. We evenly split the frequency $\lambda$ into ten buckets and calculate the normalized f($\lambda$).}
  \label{fig:GFT analyze}
\end{figure}

Generally,  the user behaviors in a system can be modeled to a signal graph, where users and items are nodes, edges represent interactions on the graph, and user and node features serve as signals on the graph. As introduced in Sec.2.3, with the help of the graph Fourier transform, we can analyze signals from a frequency perspective. To explore the characteristics of positive/negative feedback on the graph, we conduct experimental analysis in the frequency domain. Specifically, we first employ the classical Matrix Factorization \cite{FSVD} and learn one-dimensional embeddings for users and items as the signals of the graph.  Then we construct two graphs that only contain positive edges and negative edges respectively, denoted as $G^+$ and $G^-$.
Last, we apply graph Fourier transform to the $ G^+$ and $G^-$.  We conduct this experiment on ml-100k dataset and show the distribution of frequency in Fig.\ref{fig:GFT analyze}. 
By comparing Fig~\ref{fig:GFT analyze} (a) and Fig~\ref{fig:GFT analyze} (b), we can observe that the signal of $G^+$ is prominent in low frequencies, while $G^-$ is prominent in high-frequency signals. Recall that, the low frequencies represent the smooth signals, and high frequencies represent unsmooth signals. In the temporal domain, smoothness can be defined as the magnitude of the signal change between timestep $t-1$ and $t$.  Similarly, \textbf{in the graph, smoothness can be measured by the difference between the nodes connected by an edge\cite{shuman2013emerging}}. The smaller the difference between nodes connected by an edge, the smoother the graph. Therefore the results in Fig~\ref{fig:GFT analyze} align with intuition. In recommendation,  we generally assume that users have higher similarity to items they like, and correspondingly, the $G^{+}$ is dominated by low-frequency signal.
On the contrary, users are dissimilar to items they dislike, and correspondingly, G$^-$ exhibits more high-frequency signals. Most existing GNNs actually act as a low-passing filter\cite{lightgcn,kipf2016semi,wu2019simplifying}, which means the GNN extracts low-frequency signals while discarding high-frequency signals. Unfortunately, most existing Graph-based recommendation models adopt the low-passing GNN to encode graphs. Thus, they may not be suitable for capturing negative feedback signals.  
% \begin{figure}[!hbtp]

% \centering
% \includegraphics[width=0.99\columnwidth]{GFT.png}
% \caption{The title is Waiting}
% \label{fig:GFT analyze}
% \end{figure}

\subsection{The Representation Degeneration Problem in Graph-based Recommendation}
\begin{figure}[!h]
  \setlength{\abovecaptionskip}{0cm} 
        \setlength{\belowcaptionskip}{0pt}
  \centering
\captionsetup[subfigure]{skip=0pt,margin=0pt}
  \subfigure[NCF]{
    \includegraphics[width=0.32\linewidth]{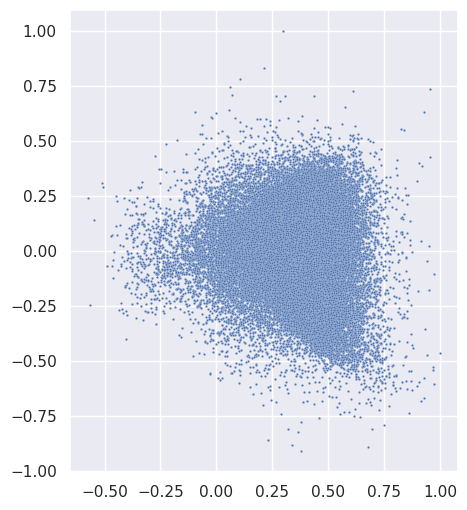}
  }\hspace{-7mm}
  \hfill
  \subfigure[GCN]{
    \includegraphics[width=0.32\linewidth]{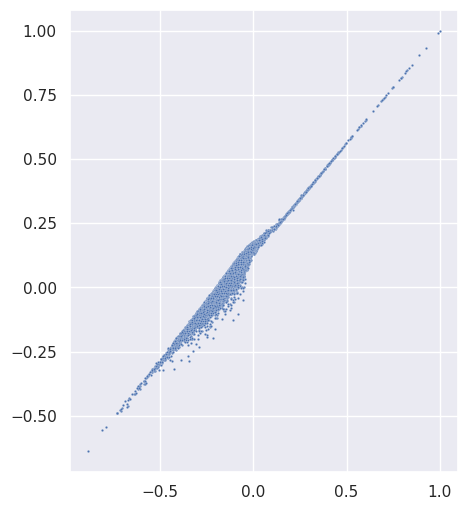}
  }
 \hfill\hspace{-7mm}
  \subfigure[Singular values]{
    \includegraphics[width=0.32\linewidth]{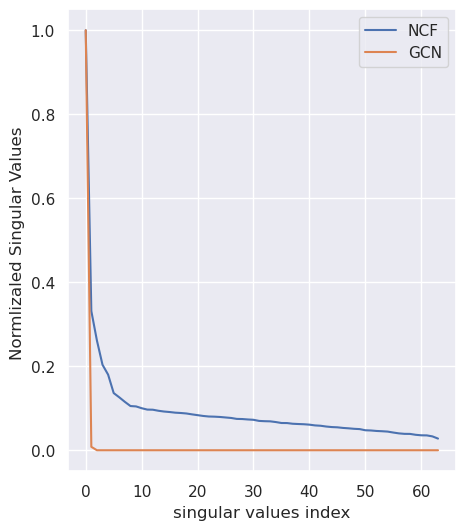}
  }
  \caption{The visualized embedding and singular values. (a) is the embedding learned by NCF.  (b) is the embedding learned by GCN. (c) is the normalized singular values of GCN and NCF.}
  \label{fig:embedding analyze}
\end{figure}

Recently, graph neural networks have achieved great success in recommendation systems \cite{lightgcn,wu2022graph,ying2018graph}. %还要加引用
However, GNNs are known as the over-smoothing effect \cite{cai2020note,elinas2022addressing}, which means the node feature tends to be similar during the GNN training, especially in deep graph neural networks. The over-smoothing effect sometimes will cause the representation degeneration problem. In this issue, the embeddings shrink to a small piece of space, thus losing the expressive power. To explore whether the graph-based recommendation will suffer from this problem, we train a graph model (i.e. GCN\cite{kipf2016semi}) and a non-graph model (i.e. NCF\cite{he2017neural}) on Amazon ArtsCrafts review dataset, respectively. We project the learned embedding matrix into 2D by SVD and visualize it. The results are shown in Fig.~\ref{fig:embedding analyze}. From this figure, we can find that the embedding learned by NCF is more uniform than that learned by GCN. Most node embeddings learned by GCN fall into a narrow zone. Furthermore, we analyze the distribution of singular values of embedding matrices and show the results in Fig.\ref{fig:embedding analyze}. We can observe a rapid decline in the singular values of GCN, indicating that the learned embeddings are low-rank and have less expressive power.
\section{Method}
Based on the above experimental observations, in this paper, we propose a novel graph model called DFGNN to solve the issues in the sign-aware graph recommendation. Our DFGNN contains two components: (1) dual frequency graph filter and (2) graph regularization. The dual-frequency graph filter aims to both capture the high-frequency signal and the low-frequency in the negative feedback and positive feedback respectively. Furthermore, the graph regularization module is adopted to alleviate the representation degeneration problem. In the following, we will make a detailed introduction to our DFGNN.

\subsection{Dual Frequency Graph Filter}
The negative feedback is crucial and widely present in recommendation systems, which show what people dislike or even hate what kind of items. However, how to deal with negative feedback is not a trivial problem.  As the analysis in Sec~\ref{sec:frequency-domain anlayze}, different from the positive feedback,  the negative feedback usually acts as the high-frequency signal in the graph.  
Hence we propose a dual-frequency graph filter (DGF), which uses a low-passing graph filter (LGF) to capture the low-frequency signal in positive feedback and adopts a high-passing graph filter (HGF) to capture the high-frequency signal in negative feedback. 

\subsubsection{Graph Filter}
The convolution operation is generally adopted as a filter to extract useful signals. The graph convolution can be defined as:
\begin{equation}
\label{eq:graph Convolution}
\begin{aligned}
x*g=& \mathcal{F}^{-1}(\mathcal{F}(x)*\mathcal{F}(g))= \mathcal{F}^{-1}(f(\lambda)g(\lambda))\\=&   E((E^Tg) \dot (E^Tx))= E\,g(\lambda)\,E^T\,x,
\end{aligned}
\end{equation}
where $g(\lambda)$  is the filter kernel function with respect to the eigenvalues $\lambda$ of the Laplacian matrix. Generally, for a low-pass filter, $g(\lambda)$ is high at low $\lambda$ but low at high $\lambda$. For a high-pass filter, it is the opposite. $(g(\lambda)$ is high at high $\lambda$ and low at low $\lambda$.

\subsubsection{Low-passing Graph Filter}
After defining the graph filter, we can design the low-passing filter. Luckily, most existing GNNs act as low-passing graph filters. In this work, we adopt the classical GCNs \cite{kipf2016semi} as our low-passing graph filter. The GCNs can be formulated as:
\begin{equation}
H^{l+1}=\sigma(\hat{A}H^lW^l), H^0=X,
\end{equation}
where $W^l\in R^{d1*d2}$ is the trainable parameters of the network, $H^{l}\in R^{N*d1}$ is the node representation after $l$ layer convolution.  $\hat{A}=\tilde{D}^{-\frac{1}{2}} \tilde{A} \tilde{D}^{\-\frac{1}{2}}$ is a designed matrix, where $\tilde{A}=A+I$  are the augmented adjacent matrix  with adding self-loops and $\tilde{D}=D+I$ are the degree matrix of $\tilde{A}$.

\subsubsection{High-passing Graph Filter}
\begin{figure}[!h]
\setlength{\abovecaptionskip}{0cm} 
        \setlength{\belowcaptionskip}{0pt}
\centering
\includegraphics[width=0.99\columnwidth]{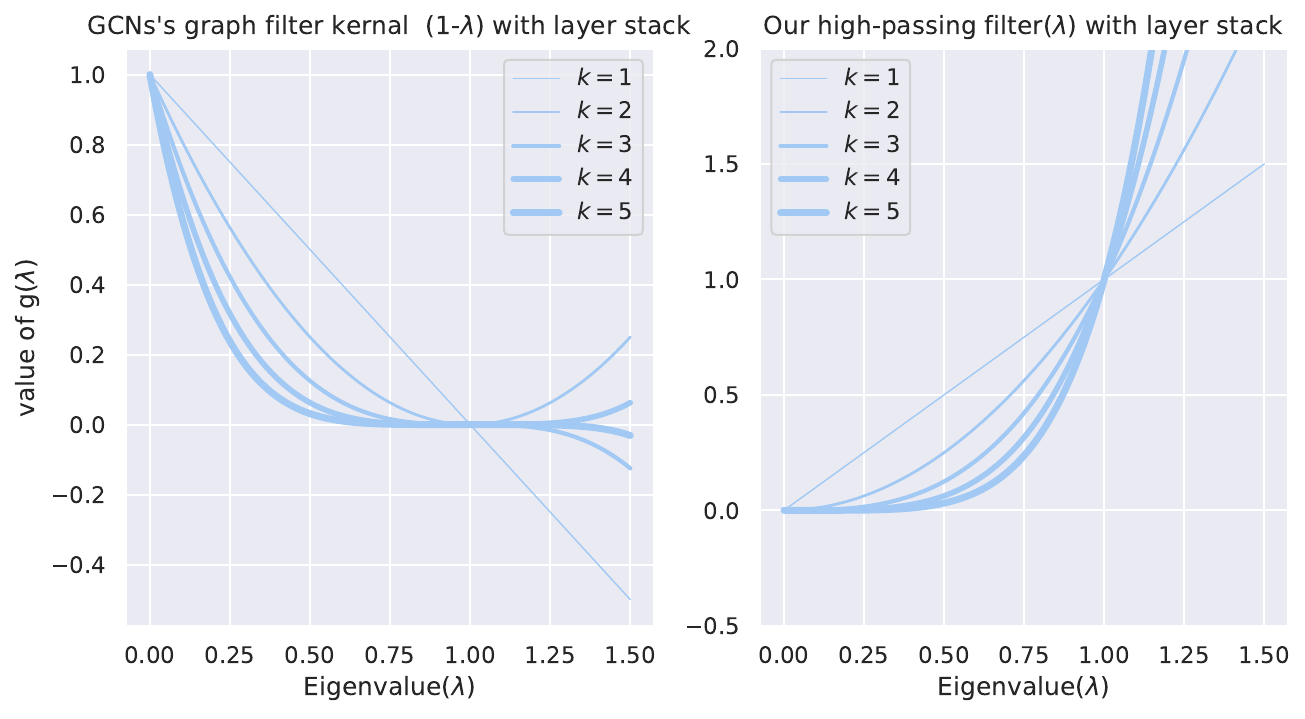}
\caption{The graph filter kernel function of LGF and HGF with layer stack. The left is the kernel (1-$\lambda$) and the right is$\lambda$ }
\label{fig:the frequency reponse}
\end{figure}
The Eq.~(\ref{eq:graph Convolution}) guides us on how to design a high-passing graph filter. However, it still is not a trivial task, as the eigenvectors $E$ are difficult to compute. Hence, we design a high-passing filter based on the existing low-passing filter (i.e., GCNs\cite{kipf2016semi}).  To achieve it,
We first hand on \textbf{why GCNs are low-passing filter}.  The normalized  Laplacian matrix \cite{shuman2013emerging} is defined  as  $L=I-D^{-\frac{1}{2}}AD^{-\frac{1}{2}}$.  Thus, the matrix $\hat{A}$ can be written as $\hat{A}=I-(I-\tilde{D}^{-\frac{1}{2}}\tilde{A}\tilde{D}^{-\frac{1}{2}})=I-\tilde{L}$. As $\tilde{L}=\tilde{E}\tilde{\Lambda}\tilde{E}^T$, the $\hat{A}H$ can be transformed to $\hat{A}H=(I-\tilde{L})H=\tilde{E}(I-\tilde{\Lambda})\tilde{E}^TH$. Comparing with Eq.(\ref{eq:graph Convolution}), we can find that the graph kernel of one layer GCNs is $g(\lambda)=1-\tilde{\lambda}$. With stacking of $K$ layer GCNs the graph kernel is $g(\lambda)=(1-\tilde{\lambda})^{K}$.  Moreover, the eigenvalues of the normalized Laplacian matrix are $[0,2)$, and the eigenvalues of the Laplacian matrix augmented by adding a self-loops are $[0,1.5)$. This implies that GCN will amplify low-frequency signals and attenuate high-frequency signals, as shown in Fig.~\ref{fig:the frequency reponse} (a). Thus GCNs actually act as low-passing filters. 

After analyzing why GCNs is a low-passing graph filter, we design the high-passing  as :
\begin{equation}
H^{l+1}= \sigma(LHW)=\sigma(D^{-\frac{1}{2}}(D-A)D^{-\frac{1}{2}}HW), H^0=X,
\end{equation}
where $A$ and $D$ are the adjacent matrix and degree matrix without augmentation. The kernel filter function of the high-passing filter is $g(\lambda)=\lambda$. As shown in Fig.~\ref{fig:the frequency reponse} it will attenuate low-frequency signals and amplify high-frequency signals.

\subsection{Graph Encoder}
In this section, we introduce how to encode the graph information with our DGF. Generally, the user feedback in recommendation contains various noises.  Looking back to Fig.~\ref{fig:Frequency}, we can see that for positive feedback there still exists high-frequency information, which can be regarded as noise in the graph. The low-passing filter usually is utilized to denoise.  Hence, we first use LGF on the positive feedback. It has two functions: (1) capturing what users like by the positive feedback. (2) weaking the noise in the graph. It can be formulated as:

\begin{equation}
  H_{pos}^{l+1}={LGF}^{l}(H^{l},G+)
\end{equation}

After applying LGF to positive feedback, we apply HGF to negative feedback to capture what users dislike.  As analyzed before our HGF will amplify high-frequency signals. It can be formulated as:
\begin{equation}
  H_{neg}^{l+1}={HGF}^{l}(H^{l},G^-)
\end{equation}

Lastly, we fusion the information aggregated from positive feedback and negative feedback with an MLP layer, we formulate it as:
\begin{equation}
  H^{l+1}=MLP(H_{pos}^{l+1}||H_{neg}^{l+1})
\end{equation}

\subsection{Signed Graph Regularization }
As discussed before, the graph neural network in recommendation may suffer from representation degeneration problems, which means the graph embedding shrinks to a piece of embedding space and loses representative power. Intuitively, if we could make the embedding uniformly distributed in the space, this issue could be alleviated. Uniform means that the average distance of each node to all nodes should be as large as possible, so the uniform constraint loss is defined as:
\begin{equation}
\begin{aligned}
  l_{uniform}=\sum^{|U|}_{u\in U}\log \sum^{|V|}_{v\in V}e^{sim({u},v)/\tau},
\end{aligned}
\end{equation}
where $u$ and $v$ are the learned embedding of user and item, the $sim(\cdot)$ is the distance function, $\tau$ is temperature. In this work, we adopt cosine similarity to measure the distance between two nodes.

However, complete uniform distribution still will lead to a loss of representational capacity issue. Because the core of recommendation systems lies in matching items that are similar to the user's interests. This implies that representations of some embedding should be similar. Furthermore, in sign-aware recommendation, things become even more complex. Negative feedback implies that some users and items are dissimilar, and their representations in the space should be as far apart as possible.  Therefore, to simultaneously consider uniformity and similarity/dis-similarity between user and items, we designed alignment loss, we formulated it as:
\begin{equation}
\begin{aligned}
  l_{alignment}&=-\sum_{(u,v)\in \mathcal{E}^+}\log e^{sim({u},v)/\tau}+\sum_{(u,v)\in \mathcal{E}^-}\log e^{sim({u},v)/\tau}\\
  &=-\sum_{(u,v)\in \mathcal{E}^+}{sim}({u},v)/\tau+\sum_{(u,v)\in \mathcal{E}^-}{sim}({u},v)/\tau,
\end{aligned}
\end{equation}
where the $\mathcal{E}^+$ and $\mathcal{E}^-$ are positive edge sets and negative edge set respectively. In fact,   $l_{alignment}$ and $l_{uniform}$  engage in a two-player game. $l_{uniform}$ aims to make the learned embeddings more uniform, while $l_{alignment}$ aims to make the learned embeddings more extreme. Finally, the regularization loss is defined as:
\begin{equation}
\begin{aligned}
  l_{SGR}=l_{uniform}+l_{alignment}.
\end{aligned}
\end{equation}
In this work, we apply our SGR on the user and item embedding layer, while do not apply it to the encoded node representation. Because we empirically found that seting results in worse performance. It may be because the encoded representation in the top layer is highly task-dependent, and forcing the representation to be uniformly distributed may not be effective.
\subsection{Loss Function}
After getting the encoded representation of the user and item. We can predict the value of $P(u_i,v_j)$. Following previous work\cite{he2017neural,kang2018self,lightgcn}, here we directly use the inner product of $h_{u_i}$ and $h_{v_j}$ to calculate it:
\begin{equation}
\begin{aligned}
  p(u_i,v_j)=Sigmoid(h_{u_i}^T\cdot h_{v_j})
\end{aligned}
\end{equation}
In this paper, we adopt classical binary cross-entropy loss\cite{kang2018self,huang2021signed} to optimize the model. The total loss is defined as:

\begin{equation}
\begin{aligned}
  l_{total}&=l_{task}+w\cdot l_{SGR}, \\
  l_{task}&=\sum^{N}yp(u_i,v_j)+(1-y)(1-p(u_i,v_j),
\end{aligned}
\end{equation}
where $w$ is the weight of $l_{SGR}$. 

% In this work, we evaluate our DFGNN on two task: (1) recommendation ranking task (2) like/dislike recognition.
% The first task is a classical recommendation task, which aims to rank items that a user likes in the candidate item set. The second one aims to predict whether a user will give positive or negative feedback to an item. For the task one, the loss function is defined as

\section{Experiment}

\begin{table*}[!htbp]
        \setlength{\abovecaptionskip}{0cm} 
        \setlength{\belowcaptionskip}{0pt}
        %\setlength{\abovecaptionskip}{0pt}
        %\setlength{\belowcaptionskip}{0pt}
        % \renewcommand{\arraystretch}{1.2}
        % \renewcommand\tabcolsep{3.0pt}
        \caption{Results on recommendation task. All improvements are significant over baselines (t-test with p ${<}$ 0.05).}
        \label{tab:recommendation}
        \center
        \setlength{\tabcolsep}{9pt}
        \begin{tabular}{l|l|cccccc|cc}
        \toprule
        \small
        Dataset & Model & GCN & GAT & SGCN & SBGNN & SBGCL & SIGRec & DFGNN & Improve \\
        \midrule
\multirow{5} {*} {ML1M}
~ & MRR &0.5085 & 0.5096 & 0.4866 & 0.5177 & 0.4707 & \underline{0.5259} & \textbf{0.5433} & 3.31\%\\
~ & HIT@10 &0.8132 & 0.8234 & 0.8041 & 0.8279 & 0.7932 & \underline{0.8381} & \textbf{0.8576} & 2.33\%\\
~ & NDCG@10 &0.2850 & 0.2882 & 0.2696 & \underline{0.2972} & 0.2645 & 0.2948 & \textbf{0.3118} & 4.91\%\\
~ & HIT@50 &0.9524 & 0.9592 & 0.9558 & 0.9587 & 0.9527 & \underline{0.9608} & \textbf{0.9628} & 0.21\%\\
~ & NDCG@50 &0.3013 & 0.3060 & 0.2912 & 0.3153 & 0.2877 & \underline{0.3184} & \textbf{0.3311} & 3.99\%\\
\midrule
 \multirow{5} {*} {Arts}
~ & MRR &0.0291 & 0.0296 & 0.0324 & \underline{0.0441} & 0.0301 & 0.0296 & \textbf{0.0643} & 45.80\%\\
~ & HIT@10 &0.0406 & 0.0427 & 0.0482 & \underline{0.0746} & 0.0432 & 0.0409 & \textbf{0.1107} & 48.39\%\\
~ & NDCG@10 &0.0230 & 0.0234 & 0.0254 & \underline{0.0350} & 0.0236 & 0.0231 & \textbf{0.0521} & 48.86\%\\
~ & HIT@50 &0.1038 & 0.1067 & 0.1093 & \underline{0.1799} & 0.1101 & 0.1076 & \textbf{0.2284} & 26.96\%\\
~ & NDCG@50 &0.0323 & 0.0326 & 0.0342 & \underline{0.0516} & 0.0334 & 0.0328 & \textbf{0.0713} & 38.18\%\\
\midrule
%  \multirow{5} {*} {Office}
% ~ & MRR &0.0238 & \underline{0.0307} & 0.0192 & 0.0275 & 0.0238 & 0.0191 & \textbf{0.0668} & 117.59\%\\
% ~ & HIT@10 &0.0486 & \underline{0.0632} & 0.0370 & 0.0549 & 0.0500 & 0.0388 & \textbf{0.1222} & 93.35\%\\
% ~ & NDCG@10 &0.0167 & \underline{0.0224} & 0.0128 & 0.0205 & 0.0173 & 0.0130 & \textbf{0.0561} & 150.45\%\\
% ~ & HIT@50 &0.1329 & \underline{0.1618} & 0.1140 & 0.1410 & 0.1351 & 0.1134 & \textbf{0.2124} & 31.27\%\\
% ~ & NDCG@50 &0.0303 & \underline{0.0387} & 0.0250 & 0.0343 & 0.0310 & 0.0248 & \textbf{0.0714} & 84.5\%\\
% \midrule
 \multirow{5} {*} {GFood}
~ & MRR &0.0421 & 0.0400 & 0.0409 & \underline{0.0447} & 0.0372 & 0.0383 & \textbf{0.0475} & 6.26\%\\
~ & HIT@10 &0.0642 & 0.0599 & 0.0615 & \underline{0.0718} & 0.0597 & 0.0585 & \textbf{0.0756} & 5.29\%\\
~ & NDCG@10 &0.0325 & 0.0314 & 0.0328 & \underline{0.0363} & 0.0289 & 0.0306 & \textbf{0.0375} & 3.31\%\\
~ & HIT@50 &0.1434 & 0.1273 & 0.1248 & \underline{0.1510} & 0.1258 & 0.1189 & \textbf{0.1528} & 1.19\%\\
~ & NDCG@50 &0.0444 & 0.0417 & 0.0423 & \underline{0.0488} & 0.0390 & 0.0394 & \textbf{0.0493} & 1.02\%\\
\midrule
 \multirow{5} {*} {Kindle}
~ & MRR &0.0381 & 0.0436 & 0.0264 & \underline{0.0476} & 0.0299 & 0.0316 & \textbf{0.0703} & 47.69\%\\
~ & HIT@10 &0.0851 & 0.0930 & 0.0548 & \underline{0.0980} & 0.0621 & 0.0682 & \textbf{0.1569} & 60.10\%\\
~ & NDCG@10 &0.0221 & 0.0263 & 0.0135 & \underline{0.0315} & 0.0160 & 0.0170 & \textbf{0.0502} & 59.37\%\\
~ & HIT@50 &0.2285 & 0.2299 & 0.1555 & \underline{0.2452} & 0.1740 & 0.1928 & \textbf{0.3380} & 37.85\%\\
~ & NDCG@50 &0.0416 & 0.0455 & 0.0253 & \underline{0.0531} & 0.0303 & 0.0326 & \textbf{0.0796} & 49.91\%\\
\midrule
 \multirow{5} {*} {Yelp}
~ & MRR &0.0378 & 0.0420 & 0.0448 & \underline{0.0460} & 0.0422 & 0.0313 & \textbf{0.0511} & 11.09\%\\
~ & HIT@10 &0.0819 & 0.0893 & 0.0925 & \underline{0.0975} & 0.0902 & 0.0649 & \textbf{0.1112} & 14.05\%\\
~ & NDCG@10 &0.0204 & 0.0231 & 0.0250 & \underline{0.0257} & 0.0230 & 0.0164 & \textbf{0.0298} & 15.95\%\\
~ & HIT@50 &0.2189 & 0.2382 & 0.2370 & \underline{0.2590} & 0.2353 & 0.1917 & \textbf{0.2875} & 11.00\%\\
~ & NDCG@50 &0.0377 & 0.0432 & 0.0442 & \underline{0.0476} & 0.0423 & 0.0317 & \textbf{0.0543} & 14.08\%\\

\bottomrule
\end{tabular}
\label{tab:ranking}
\end{table*}

In this section, we aim to answer the following questions:
 \textbf{(1) RQ1:}How does DFGNN perform compared with other state-of-the-art (SOTA) baselines on recommendation  task? (Sec.~\ref{sec:Recommendation Results})  \textbf{(2) RQ2:} How does DFGSN perform compared with other state-of-the-art (SOTA) baselines on feedback type recognition task? (Sec.~\ref{sec:Feedback Type Recognition}) 
    \textbf{(3) RQ3:}  The proposed two modules in DFGNN indeed work? (Sec.~\ref{sec:ablation}) 
    \textbf{(4) RQ4:} How does our signed graph regulation loss help  in the training? (Sec.~\ref{sec:uniform}) 

\subsection{Experimental Setting}
\noindent
\textbf{Datasets} In this paper, to evaluate the performance of the proposed DFGNN, we conduct comprehensive experiments on classical \textbf{ML1M, different categories of Amazon Review datasets, and Yelp }dataset. They are all real-world rating datasets. each instance was assigned a rating from 1 to 5. we regard the ratings that are below 3 as negative feedback and the ratings that are higher than 3 as positive feedback. Due to the space limit, we detailly introduce and analyze them in the Appendix\ref{app:dataset}.

\noindent
\textbf{Evaluation Protocols.}
We evaluate the proposed model on two tasks called the recommendation ranking task and feedback type recognition task.  For the recommendation ranking task, we adopt widely accepted ranking metrics\cite{kang2018self,wang2019neural}:
top-K hit rate (HIT@k), top-K (NDCG@k), and MRR, with $k=\{10 and 50\}$.  Feedback type recognition task can be regarded as a binary classification task. Considering the imbalance between positive and negative samples(as shown in Table.\ref{tab:dataset}), here we use classical AUC and F1-Macro as evaluation metrics.  It is necessary to note that we do not adopt accuracy, precision, recall, etc., as metrics, since they are not applicable to imbalanced datasets. For each task, we report the mean results with five random runs. 

\noindent
\textbf{Implement Details.} We carefully tune all the baselines and compare them fairly. Due to the space limit, we show the details in Appendix.~\ref{app:Implement details}.

 \subsection{Baselines}
 % To validate our IDP, we conduct a comprehensive comparison. Specifically, we compare our method with the following baselines:(1)\textbf{ GRU4Rec}\cite{hidasi2016session} (2)\textbf{ SASRec} \cite{kang2018self}, (3)\textbf{ BERT4Rec} \cite{sun2019BERT4Rec}, (4)\textbf{ FDSA} \cite{zhang2019feature}, (5)\textbf{Recformer}\cite{li2023text}, and (6)\textbf{UniSRec}$_{ID+t}$ \cite{hou2022towards}. Please refer to the \textbf{Appendix.\ref{Baselines}} for a detailed description of baselines.

\label{Baselines}
  To validate the performance of DFGNN, we compare our method with classical and state-of-the-art unsigned/signed graph neural networks: (1)\textbf{ GCN}\cite{kipf2016semi}, (2)\textbf{ GAT} \cite{velivckovic2017graph}, (3) \textbf{SGCN}\cite{derr2018signed} (4)\textbf{SBGNN} \cite{huang2021signed}, (5)\textbf{ SBGCL} \cite{zhang2023contrastive} (6) \textbf{SIGRec}\cite{huang2023negative}.  
  For unsigned GNNs, we only retain the positive feedback edge.
  We implement our method by PyG\footnote{https://pytorch-geometric.readthedocs.io/en/latest/index.html}, which is a popular graph learning framework. For baseline GCN, GAT, and SGCN we directly use the implementation of PyG. For baseline SBGNN and SBGCL, we directly use the author's released codes. For SIGRec as there is no available open-sourced code, we implement it by ourselves. They are detailly introduced in Appendix~\ref{Baselines}
    % \textbf{ GRU4Rec} \cite{hidasi2016session}, which adopts GRU as  user sequence encoder for sequential recommenadtion. 
    %  \textbf{ SASRec} \cite{kang2018self}, which is a classical sequential recommendation model that introduces self-attention in behavior modeling. 
    %  \textbf{ BERT4Rec} \cite{sun2019BERT4Rec}, which adopts masked item prediction as an optimizing object.
    %  \textbf{ FDSA} \cite{zhang2019feature}, which adopts self-attentive networks to model item and feature transition. In this paper, we adopt item descriptions as item features. 
    %  \textbf{UniSRec}$_{ID+t}$ \cite{hou2022towards}, UniSRec is a SOTA text-based pre-train model. In the pre-traning stage, UniSRec represents items with text embedding encoded by a pre-trained language model and pre-train recommendation model based on text embedding. In the downstream domain, UniSRec fine-tunes pre-trained model by the designed MoE-enhanced adopter. UniSRec$_{ID+t}$ is the final version of UniSRec, which add ID embedding in downstream domains.  

  \subsection{Performance on Recommendation}
  \label{sec:Recommendation Results}
  We show the results of our DFGNN in Table~\ref{tab:ranking} on the recommendation task, which aims to recommend items to users from a set of candidates. From the table, we can see that:

(1) Compared with the unsigned graph neural network (i.e., GCN and GAT), the signed graph neural network generally outperforms them across all datasets.  It demonstrates the importance of negative feedback. The signed graph neural network both considers the positive feedback and negative feedback information in the graph, thus achieving a better performance.

  (2) We note that the SGCN generally performs worse than  SBGCL. SBGCL is specifically designed for bipartite graphs. It extends the balance theory tp the butterfly structure in bipartite graphs. It matches the graph structure characteristics in recommendation, which are a user-item bipartite graphs.  In contrast, SGCN is based on the balance theory with a triangle structure, which may lead to its suboptimal performance.

  (3) Compared with all baselines, our DFGNN significantly outperforms them on all datasets, even with improvements of over 50\% observed in some datasets.
This can be attributed to our special design for handling negative feedback.  Traditional GNNs are designed for homogeneous graphs and rely on the homophily assumption, which assumes that nodes connected by edges are similar. While in the sign-aware recommendation, users and items nodes connected by negative feedback intuitively indicate dissimilarity. Our analysis of negative signals from a frequency view also evaluates it.  Based on this, we design dual-frequency graph filters and signed graph regularization, which can break the homophily assumption, and thus better capture the signal from negative feedback.
  
      \begin{table*}[htp]
      \setlength{\abovecaptionskip}{0cm} 
        \setlength{\belowcaptionskip}{0pt}
        %\setlength{\abovecaptionskip}{0pt}
        %\setlength{\belowcaptionskip}{0pt}
        % \renewcommand{\arraystretch}{1.2}
        % \renewcommand\tabcolsep{3.0pt}
    \caption{Results on feedback type recognition task. All improvements are significant over baselines (t-test with p ${<}$ 0.05).}
    \label{tab:negative prediction}
    \center
    \setlength{\tabcolsep}{9pt}
    \begin{tabular}{l|l|cccccc|cc}
    \toprule
    \small
    Dataset & Model & GCN & GAT & SGCN & SBGNN & SBGCL & SIGRec & DFGNN & Impro \\
    \midrule
 \multirow{2} {*} {ML1M}
~ & AUC &0.8802 & 0.8838 & 0.8679 & \underline{0.8877} & 0.8861 & 0.8665 & \textbf{0.8925} & 0.54\%\\
~ & F1-Macro &0.7767 & 0.7791 & 0.7658 & 0.7800 & \underline{0.7845} & 0.7411 & \textbf{0.7902} & 0.73\%\\
\midrule
 \multirow{2} {*} {Arts}
~ & AUC &0.7366 & 0.7303 & 0.7337 & \underline{0.7377} & 0.7324 & 0.7294 & \textbf{0.7577} & 2.71\%\\
~ & F1-Macro &0.5904 & 0.5579 & 0.5777 & 0.6001 & \underline{0.6035} & 0.5662 & \textbf{0.6095} & 0.99\%\\
\midrule
%  \multirow{2} {*} {Office}
% ~ & AUC &0.7390 & 0.7358 & 0.7329 & 0.7328 & \underline{0.7394} & 0.7370 & \textbf{0.7622} & 3.08\%\\
% ~ & F1_Macro &0.5925 & 0.5909 & 0.5934 & 0.5776 & \underline{0.6041} & 0.5989 & \textbf{0.6072} & 0.51\%\\
% \midrule
 \multirow{2} {*} {Kindle}
~ & AUC &0.8591 & \underline{0.8630} & 0.8389 & 0.8324 & 0.8352 & 0.8606 & \textbf{0.8691} & 0.71\%\\
~ & F1-Macro &0.6318 & 0.6133 & 0.6500 & 0.6538 & \underline{0.6632} & 0.6250 & \textbf{0.6892} & 3.92\%\\
\midrule
 \multirow{2} {*} {GFood}
~ & AUC &\underline{0.7527} & 0.7057 & 0.7250 & 0.7371 & 0.7402 & 0.7484 & \textbf{0.7561} & 0.45\%\\
~ & F1-Macro &0.6043 & 0.5610 & 0.5979 & \underline{0.6047} & 0.5751 & 0.5970 & \textbf{0.6144} & 1.6\%\\
\midrule
 \multirow{2} {*} {Yelp}
~ & AUC &0.7705 & 0.7770 & 0.7715 & 0.7572 & \underline{0.7777} & 0.7725 & \textbf{0.7907} & 1.67\%\\
~ & F1-Macro &0.6017 & 0.6633 & \textbf{0.6841} & 0.6636 & 0.6605 & 0.5860 & \underline{0.6713} & -\\
\bottomrule
\end{tabular}
\end{table*}

  \subsection{Performance on Feedback Type Recognition}

\label{sec:Feedback Type Recognition}
   \begin{figure}[!hbp]
   \setlength{\abovecaptionskip}{0cm} 
        \setlength{\belowcaptionskip}{0pt}
\centering
\includegraphics[width=0.99\columnwidth]{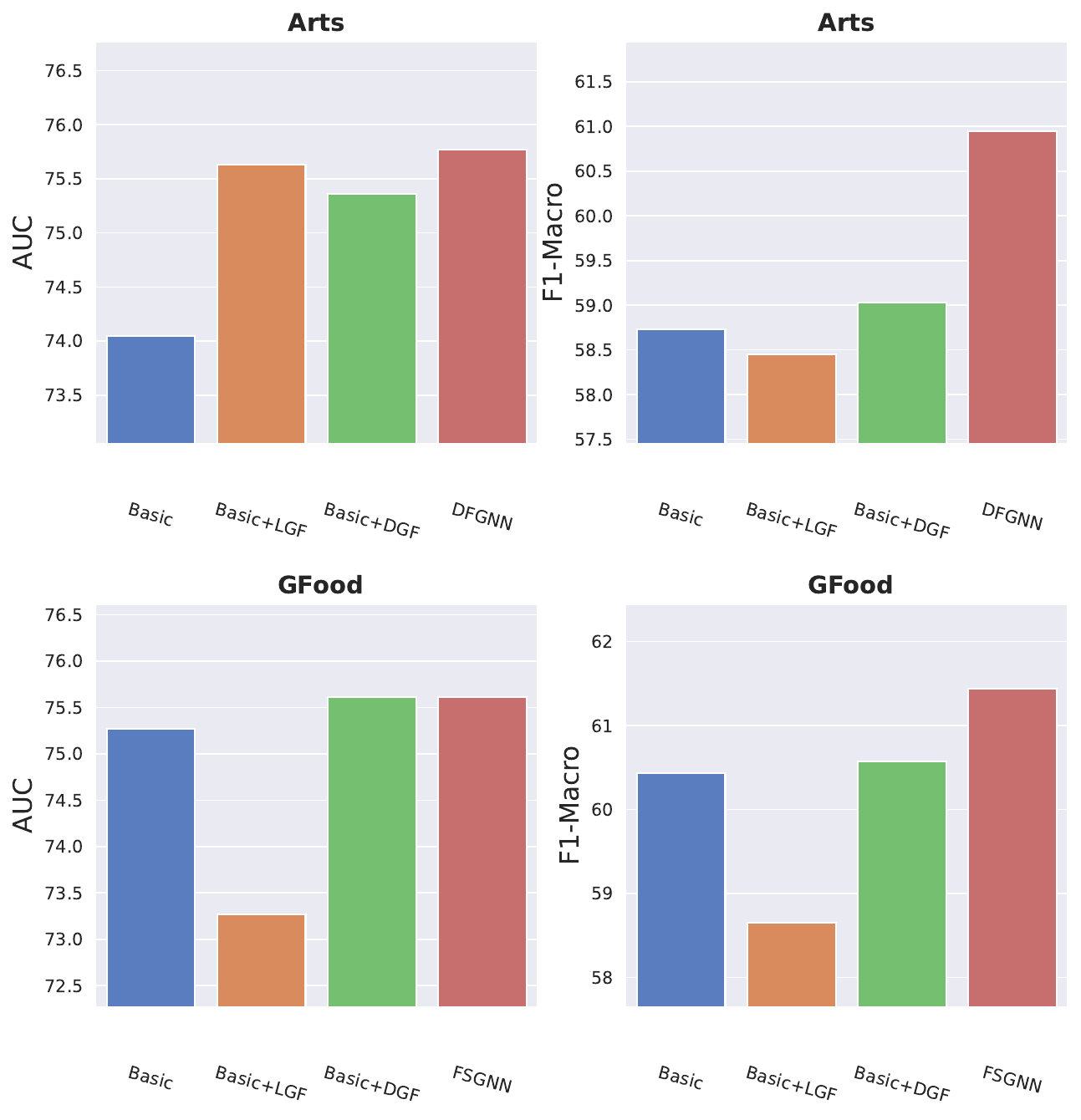}
\caption{Ablation study of feedback type recognition task on Arts and GFood dataset}
\label{fig:ablation on Reg}
\end{figure}
\begin{figure}[!h]
\setlength{\abovecaptionskip}{0cm} 
        \setlength{\belowcaptionskip}{0pt}
\centering
\includegraphics[width=0.99\columnwidth]{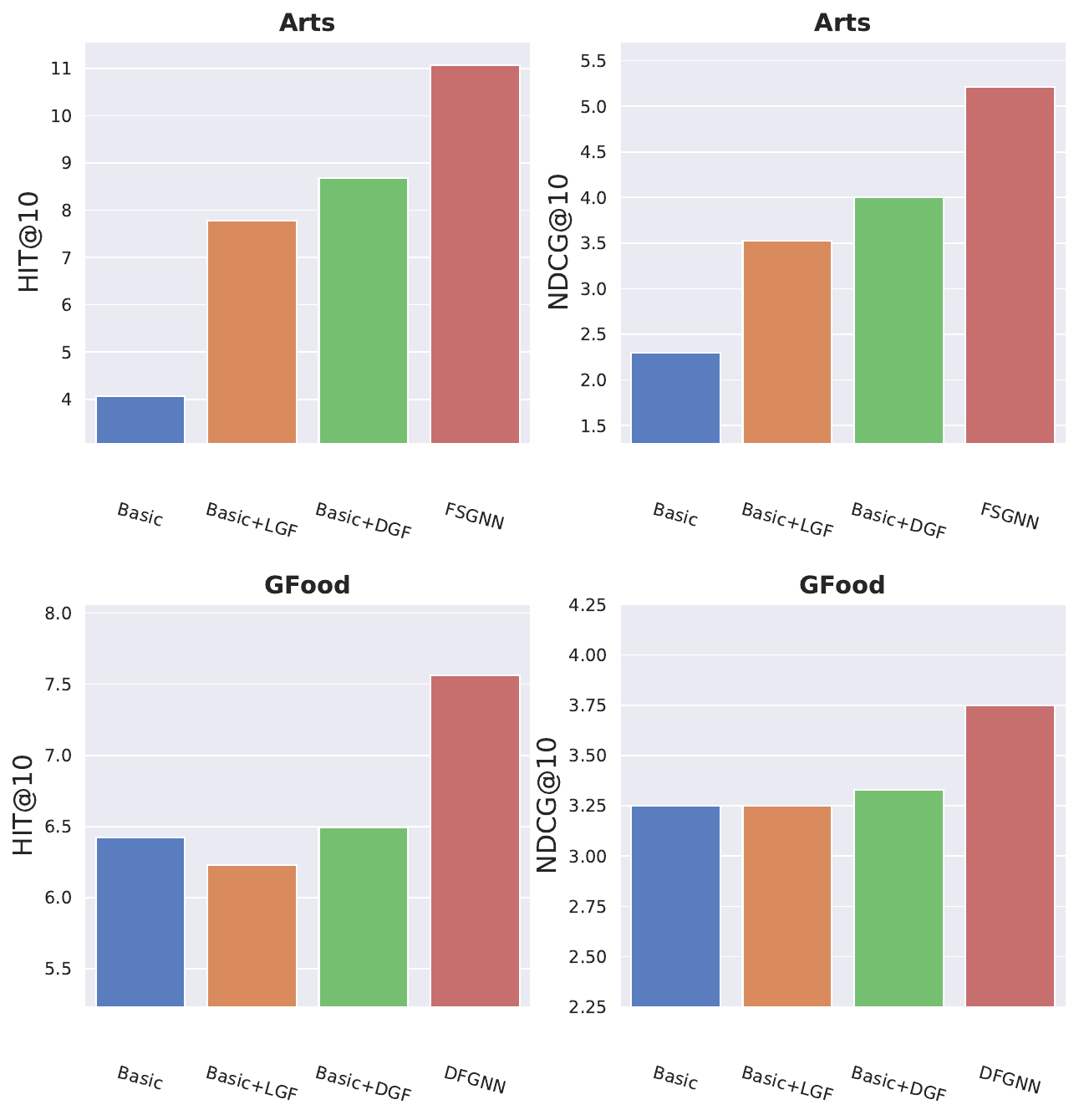}
\caption{Ablation study of recommendation task on Arts and GFood dataset}
\label{fig:ablation on Rec}
\end{figure}
This section evaluates our DFGNN on the feedback type recognition task. Given a user and item pair that the user has interacted with, this task aims to predict whether the user will give positive/negative feedback. This task focuses more on identifying what users dislike, as recommending items that users find displeasing can significantly harm the user experience.
We report the experiment results in Table~\ref{tab:negative prediction}. From this table, we have:

\begin{figure*}[!ht]
\centering
  \setlength{\abovecaptionskip}{0cm} 
        \setlength{\belowcaptionskip}{0pt}
\captionsetup[subfigure]{skip=0pt,margin=0pt}
  \subfigure[GCN on FTR task]{
    \includegraphics[width=0.24\linewidth]{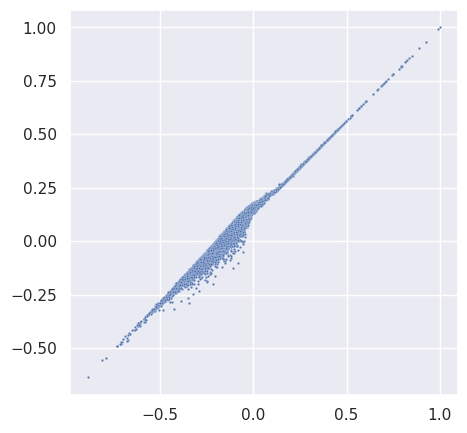}
  }\hspace{-7mm}
  \hfill
  \subfigure[DFGNN on FTR task]{
    \includegraphics[width=0.24\linewidth]{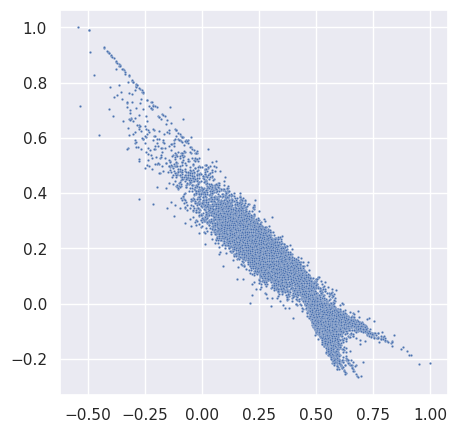}
  }
 \hfill\hspace{-7mm}
  \subfigure[GCN on RS task]{
    \includegraphics[width=0.24\linewidth]{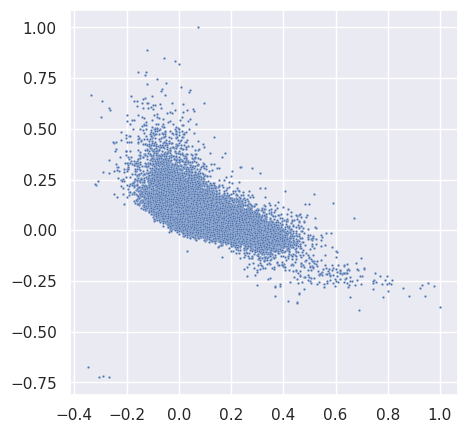}
  }\hfill\hspace{-7mm}
   \subfigure[DFGNN on RS task]{
    \includegraphics[width=0.24\linewidth]{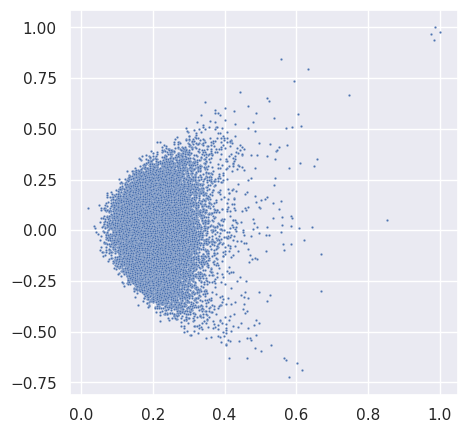}
  }
  \caption{The visualized embedding of GCN and FSGNN on two tasks. FTR means Feedback Type Recognition task, and RS means Recommendation task }.
\label{fig:uniform_analysis_v}

\end{figure*}

 (1) We note that compared to the unsigned graph models, the signed graph neural networks perform almost entirely better than them.  This may be attributed to the positive/negative feedback prediction is a harder task.  In fact, users only interact with an item when they have a certain level of interest in it. Negative feedback may be provided later after that. Compared to selecting items that users like from random negative items (task in recommendation), this task is more difficult.
 In this scenario, a lack of precise modeling of negative feedback signals may limit the improvement in prediction accuracy. 
 
(2) Compared to both signed graph neural networks and unsigned graph neural networks, our DFGNN still outperforms all of them except F1-Macro in the Yelp dataset. As analyzed before, positive/negative recognition is not trivial. Though edges constructed based on co-negative relationships (e.g., $v_1$, $v_2$ are both disliked by $u_1$) in the SBGNN and SBGCL  intuitively can utilize low-passing filters to capture negative feedback information. They still struggle to capture high-passing signals between nodes with opposing relationships (e.g., \(u_1\) dislikes \(v_1\)). With the help of our dual-frequency graph filters, our model can capture the high-frequency signal in the negative feedback. Furthermore, the signed graph regularization further alleviates the representation degeneration problem and helps the graph filters work better.

\subsection{Ablation Study}
\label{sec:ablation}

In this section, we conduct ablation experiments to evaluate the effectiveness of the proposed module in our DFGNN. We transform our model into three versions (1) Basic is the basic version of DFGNN, which only contains the positive edge and low-passing GNN  (2) Basic + LGF is the version that applies low-passing graph filter on negative edges. (3) Basic + DGF, which adopts our DGF to encode the signed graph  (4) DFGNN is the full version of our model, which both adopts our SGR and DGF to the signed graph.  We conduct the ablation experiment on the recommendation ranking and feedback type recognition task. And show the results in Fig.~\ref{fig:ablation on Reg} and  Fig.~\ref{fig:ablation on Rec}. From the figures, we can find that the proposed DGF and SGR can indeed improve the performance of two tasks. The DGF both captures the high-frequency and low-frequency signals in two types of feedback, based on this the SGR alleviates the over-smoothing problem.   
Besides, We note that the Basic model generally outperforms the Basic+LGF on the feedback type prediction task. However, in the recommendation task, the opposite is observed. This is because the negative feedback still reflects users' interest (similarity), as they interact with rather than ignore them. 
So it may benefit the recommendation task.

% which uses a low-passing filter to model negative feedback. This implies inappropriate modeling negative feedback even will harm the model's performance.

\begin{figure}[!h]
\centering
  \setlength{\abovecaptionskip}{0cm} 
        \setlength{\belowcaptionskip}{0pt}
\captionsetup[subfigure]{skip=0pt,margin=0pt}
  \subfigure[GCN on FTR task]{
    \includegraphics[width=0.32\linewidth]{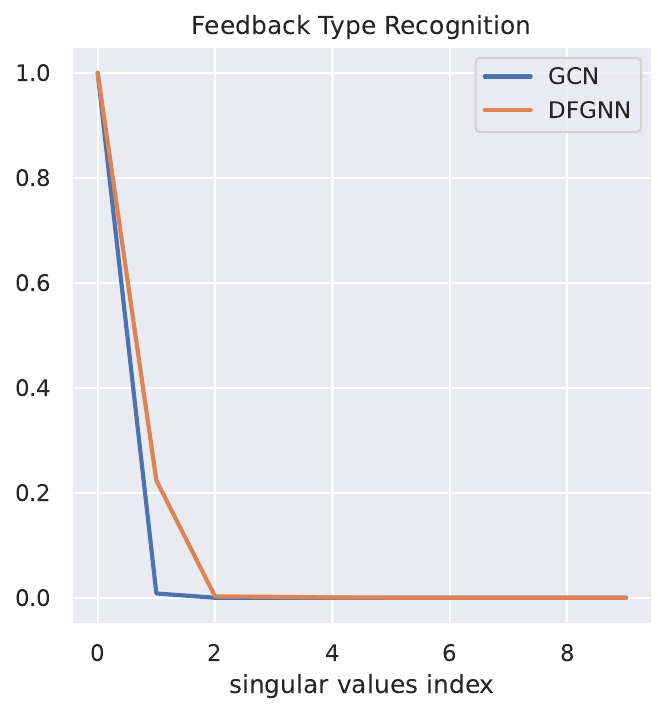}
  }\hspace{-7mm}
  \hfill
  \subfigure[DFGNN on FTR task]{
    \includegraphics[width=0.32\linewidth]{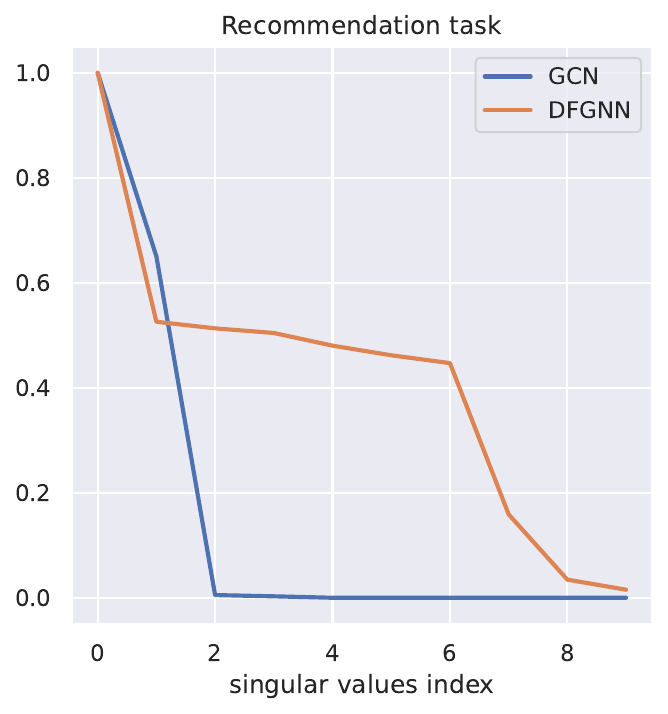}
  }
 \hfill\hspace{-7mm}
  \subfigure[Uniform of GCN and DFGNN]{
    \includegraphics[width=0.32\linewidth]{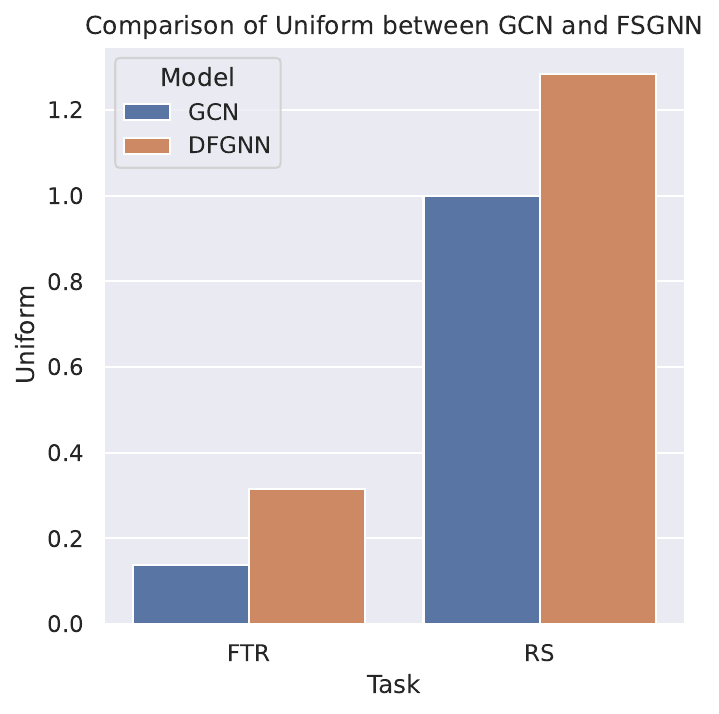}
  }
  \caption{The Singular values and the uniform (average distance between all nodes) of GCN and DFGNN on two tasks}.
\label{fig:uniform_analysis_m}
\end{figure}
% 介绍GNN
% GNN的是low-passing
% 
\vspace{-0.6cm}
\subsection{Uniform Analysis}
\label{sec:uniform}

In this paper, based on the observation of the representation degenerative problem, we propose SGR to alleviate it. To evaluate our SGR and further understand how our SGR improves performance: (1) we visualize the learned node embedding, and (2) we give out the normalized singular values of the embedding encoded by GNNs and the average distance between nodes. Specifically, we project the learned embedding to 2D by SVD for visualization. We use the average L2 distance of normalized embedding to measure uniform. The larger the value of distance, the more uniform the learned embedding is.  
We run the DFGNN and GCN on the Arts dataset. The results are shown in Fig.~\ref{fig:uniform_analysis_v} and Fig.~\ref{fig:uniform_analysis_m}. By observing the Fig.~\ref{fig:uniform_analysis_v} we can find that the DFGNN learned embedding is uniform to that of GCN. Secondly, the embedding learned by feedback type prediction task is more narrow. This implies the feedback type prediction task is a more difficult task.  Furthermore, by observing Fig.~\ref{fig:uniform_analysis_m}, We can see singular values of the embeddings learned by our DFGNN decrease more slowly. The distances of the learned embeddings are also larger. This indicates the designed SGR  indeed can help learned embedding to be more uniform. Thus, it can improve the performance of our model.

\section{Related Work}
\noindent\textbf{GNN and Graph-based Recommendation} Graph Neural networks(GNN) are studied to model graph data and have achieved great success. GCNs\cite{bruna2013spectral,defferrard2016convolutional,kipf2016semi,wu2019simplifying,gao2019graph} firstly extend convolution operations to graph data. Further, GraphSage, GAT, GIN et.al\cite{hamilton2017inductive,velivckovic2017graph,xu2018powerful} abstract GNN to a message-passing schema and propose various information aggregation strategies. Recommender system aims to recommend appropriate items to users by utilizing user-item interaction information. Recently, recommendation methods based on deep learning have achieved great success\cite{LUO,DSSM,covington2016deep,xu2023artificial,he2017neural}.
Borrowing the promising performance, GNNs are introduced to explore high-order interaction information contained in recommendations. Pinsage\cite{ying2018graph} firstly applies graph neural network to the recommendation.  NGCF\cite{wang2019neural} extends collaborative filtering to graph. LightGCN \cite{lightgcn} removes the non-linear layer between  GCN sublayer to alleviate the over-parameter issues. despite the success of graph-based recommendations, they are merely used to model positive feedback. How to model negative feedback in graph-based recommendation remains underexplored. In this work we focus on modeling negative feedback in graph-based recommendation.

\noindent\textbf{Signed graph neural network} 
In the signed graph, there are two opposite edges (e.g. reject/accept, trustworthy/trustworthy).  Signed graph neural networks are proposed to deal with such graph structures\cite{jung2020signed,kim2018side,yuan2017sne,li2020learning,huang2019signed}. SGCN\cite{derr2018signed} first introduces the GCN to the signed graph by utilizing the balance theory. SBGNN\cite{huang2021signed} extends the balance theory to the bipartite graph. SBGCL\cite{zhang2023contrastive} modified the graph contrastive learning to the customized signed graph.   However, most of them do not focus on the recommendation scenario and adopt classical graph neural networks as encoders, which heavily rely on homophily assumption. In this paper, we find this assumption may not be invalid for recommendation.
% \vspace{-2em}
\section{Conclusions}
In this paper,  we first essentially analyze the negative and positive feedback from the graph signal frequency perspective. Based on our experimental observations, we propose a novel Dual-frequency Graph Neural Network (DFGNN), which models positive/negative feedback with a designed dual-frequency filter and alleviates representation degeneration problem with a signed-graph regularization loss. Extensive experiments validate the power of our FDGNN.
% \section{Method}
\bibliographystyle{ACM-Reference-Format}
\bibliography{sample-base}


\begin{thebibliography}{35}

%%% the Web address.

\ifx \showCODEN    \undefined \def \showCODEN     #1{\unskip}     \fi
\ifx \showDOI      \undefined \def \showDOI       #1{#1}\fi
\ifx \showISBNx    \undefined \def \showISBNx     #1{\unskip}     \fi
\ifx \showISBNxiii \undefined \def \showISBNxiii  #1{\unskip}     \fi
\ifx \showISSN     \undefined \def \showISSN      #1{\unskip}     \fi
\ifx \showLCCN     \undefined \def \showLCCN      #1{\unskip}     \fi
\ifx \shownote     \undefined \def \shownote      #1{#1}          \fi
\ifx \showarticletitle \undefined \def \showarticletitle #1{#1}   \fi
\ifx \showURL      \undefined \def \showURL       {\relax}        \fi
% invisible to TeX
\providecommand\bibfield[2]{#2}
\providecommand\bibinfo[2]{#2}
\providecommand\natexlab[1]{#1}
\providecommand\showeprint[2][]{arXiv:#2}

\bibitem[Bruna et~al\mbox{.}(2013)]%
        {bruna2013spectral}
\bibfield{author}{\bibinfo{person}{Joan Bruna}, \bibinfo{person}{Wojciech Zaremba}, \bibinfo{person}{Arthur Szlam}, {and} \bibinfo{person}{Yann LeCun}.} \bibinfo{year}{2013}\natexlab{}.
\newblock \showarticletitle{Spectral networks and locally connected networks on graphs}.
\newblock \bibinfo{journal}{\emph{arXiv preprint arXiv:1312.6203}} (\bibinfo{year}{2013}).
\newblock


\bibitem[Cai and Wang(2020)]%
        {cai2020note}
\bibfield{author}{\bibinfo{person}{Chen Cai} {and} \bibinfo{person}{Yusu Wang}.} \bibinfo{year}{2020}\natexlab{}.
\newblock \showarticletitle{A note on over-smoothing for graph neural networks}.
\newblock \bibinfo{journal}{\emph{arXiv preprint arXiv:2006.13318}} (\bibinfo{year}{2020}).
\newblock


\bibitem[Covington et~al\mbox{.}(2016)]%
        {covington2016deep}
\bibfield{author}{\bibinfo{person}{Paul Covington}, \bibinfo{person}{Jay Adams}, {and} \bibinfo{person}{Emre Sargin}.} \bibinfo{year}{2016}\natexlab{}.
\newblock \showarticletitle{Deep neural networks for youtube recommendations}. In \bibinfo{booktitle}{\emph{Proceedings of the 10th ACM conference on recommender systems}}. \bibinfo{pages}{191--198}.
\newblock


\bibitem[Defferrard et~al\mbox{.}(2016)]%
        {defferrard2016convolutional}
\bibfield{author}{\bibinfo{person}{Micha{\"e}l Defferrard}, \bibinfo{person}{Xavier Bresson}, {and} \bibinfo{person}{Pierre Vandergheynst}.} \bibinfo{year}{2016}\natexlab{}.
\newblock \showarticletitle{Convolutional neural networks on graphs with fast localized spectral filtering}.
\newblock \bibinfo{journal}{\emph{Advances in neural information processing systems}}  \bibinfo{volume}{29} (\bibinfo{year}{2016}).
\newblock


\bibitem[Derr et~al\mbox{.}(2018)]%
        {derr2018signed}
\bibfield{author}{\bibinfo{person}{Tyler Derr}, \bibinfo{person}{Yao Ma}, {and} \bibinfo{person}{Jiliang Tang}.} \bibinfo{year}{2018}\natexlab{}.
\newblock \showarticletitle{Signed graph convolutional networks}. In \bibinfo{booktitle}{\emph{2018 IEEE International Conference on Data Mining (ICDM)}}. IEEE, \bibinfo{pages}{929--934}.
\newblock


\bibitem[Elinas and Bonilla(2022)]%
        {elinas2022addressing}
\bibfield{author}{\bibinfo{person}{Pantelis Elinas} {and} \bibinfo{person}{Edwin~V Bonilla}.} \bibinfo{year}{2022}\natexlab{}.
\newblock \showarticletitle{Addressing Over-Smoothing in Graph Neural Networks via Deep Supervision}.
\newblock \bibinfo{journal}{\emph{arXiv preprint arXiv:2202.12508}} (\bibinfo{year}{2022}).
\newblock


\bibitem[Funk({[n.\,d.]})]%
        {FSVD}
\bibfield{author}{\bibinfo{person}{Simon Funk}.} \bibinfo{year}{[n.\,d.]}\natexlab{}.
\newblock \showarticletitle{Funk’s original post}.
\newblock
\urldef\tempurl%
\url{https://sifter.org/~simon/journal/20061211.html}
\showURL{%
\tempurl}


\bibitem[Gao and Ji(2019)]%
        {gao2019graph}
\bibfield{author}{\bibinfo{person}{Hongyang Gao} {and} \bibinfo{person}{Shuiwang Ji}.} \bibinfo{year}{2019}\natexlab{}.
\newblock \showarticletitle{Graph u-nets}. In \bibinfo{booktitle}{\emph{international conference on machine learning}}. PMLR, \bibinfo{pages}{2083--2092}.
\newblock


\bibitem[Hamilton et~al\mbox{.}(2017)]%
        {hamilton2017inductive}
\bibfield{author}{\bibinfo{person}{Will Hamilton}, \bibinfo{person}{Zhitao Ying}, {and} \bibinfo{person}{Jure Leskovec}.} \bibinfo{year}{2017}\natexlab{}.
\newblock \showarticletitle{Inductive representation learning on large graphs}.
\newblock \bibinfo{journal}{\emph{Advances in neural information processing systems}}  \bibinfo{volume}{30} (\bibinfo{year}{2017}).
\newblock


\bibitem[He et~al\mbox{.}(2020)]%
        {lightgcn}
\bibfield{author}{\bibinfo{person}{Xiangnan He}, \bibinfo{person}{Kuan Deng}, \bibinfo{person}{Xiang Wang}, \bibinfo{person}{Yan Li}, \bibinfo{person}{YongDong Zhang}, {and} \bibinfo{person}{Meng Wang}.} \bibinfo{year}{2020}\natexlab{}.
\newblock \showarticletitle{LightGCN: Simplifying and Powering Graph Convolution Network for Recommendation}. In \bibinfo{booktitle}{\emph{Proceedings of the 43rd International ACM SIGIR Conference on Research and Development in Information Retrieval}} (Virtual Event, China) \emph{(\bibinfo{series}{SIGIR '20})}. \bibinfo{publisher}{Association for Computing Machinery}, \bibinfo{address}{New York, NY, USA}, \bibinfo{pages}{639–648}.
\newblock
\showISBNx{9781450380164}
\urldef\tempurl%
\url{https://doi.org/10.1145/3397271.3401063}
\showDOI{\tempurl}


\bibitem[He et~al\mbox{.}(2017)]%
        {he2017neural}
\bibfield{author}{\bibinfo{person}{Xiangnan He}, \bibinfo{person}{Lizi Liao}, \bibinfo{person}{Hanwang Zhang}, \bibinfo{person}{Liqiang Nie}, \bibinfo{person}{Xia Hu}, {and} \bibinfo{person}{Tat-Seng Chua}.} \bibinfo{year}{2017}\natexlab{}.
\newblock \showarticletitle{Neural collaborative filtering}. In \bibinfo{booktitle}{\emph{Proceedings of the 26th international conference on world wide web}}. \bibinfo{pages}{173--182}.
\newblock


\bibitem[Huang et~al\mbox{.}(2021)]%
        {huang2021signed}
\bibfield{author}{\bibinfo{person}{Junjie Huang}, \bibinfo{person}{Huawei Shen}, \bibinfo{person}{Qi Cao}, \bibinfo{person}{Shuchang Tao}, {and} \bibinfo{person}{Xueqi Cheng}.} \bibinfo{year}{2021}\natexlab{}.
\newblock \showarticletitle{Signed bipartite graph neural networks}. In \bibinfo{booktitle}{\emph{Proceedings of the 30th ACM International Conference on Information \& Knowledge Management}}. \bibinfo{pages}{740--749}.
\newblock


\bibitem[Huang et~al\mbox{.}(2019)]%
        {huang2019signed}
\bibfield{author}{\bibinfo{person}{Junjie Huang}, \bibinfo{person}{Huawei Shen}, \bibinfo{person}{Liang Hou}, {and} \bibinfo{person}{Xueqi Cheng}.} \bibinfo{year}{2019}\natexlab{}.
\newblock \showarticletitle{Signed graph attention networks}. In \bibinfo{booktitle}{\emph{Artificial Neural Networks and Machine Learning--ICANN 2019: Workshop and Special Sessions: 28th International Conference on Artificial Neural Networks, Munich, Germany, September 17--19, 2019, Proceedings 28}}. Springer, \bibinfo{pages}{566--577}.
\newblock


\bibitem[Huang et~al\mbox{.}(2023)]%
        {huang2023negative}
\bibfield{author}{\bibinfo{person}{Junjie Huang}, \bibinfo{person}{Ruobing Xie}, \bibinfo{person}{Qi Cao}, \bibinfo{person}{Huawei Shen}, \bibinfo{person}{Shaoliang Zhang}, \bibinfo{person}{Feng Xia}, {and} \bibinfo{person}{Xueqi Cheng}.} \bibinfo{year}{2023}\natexlab{}.
\newblock \showarticletitle{Negative can be positive: Signed graph neural networks for recommendation}.
\newblock \bibinfo{journal}{\emph{Information Processing \& Management}} \bibinfo{volume}{60}, \bibinfo{number}{4} (\bibinfo{year}{2023}), \bibinfo{pages}{103403}.
\newblock


\bibitem[Huang et~al\mbox{.}(2013)]%
        {DSSM}
\bibfield{author}{\bibinfo{person}{Po-Sen Huang}, \bibinfo{person}{Xiaodong He}, \bibinfo{person}{Jianfeng Gao}, \bibinfo{person}{Li Deng}, \bibinfo{person}{Alex Acero}, {and} \bibinfo{person}{Larry Heck}.} \bibinfo{year}{2013}\natexlab{}.
\newblock \showarticletitle{Learning deep structured semantic models for web search using clickthrough data} \emph{(\bibinfo{series}{CIKM '13})}. \bibinfo{publisher}{Association for Computing Machinery}, \bibinfo{address}{New York, NY, USA}, \bibinfo{pages}{2333–2338}.
\newblock
\showISBNx{9781450322638}
\urldef\tempurl%
\url{https://doi.org/10.1145/2505515.2505665}
\showDOI{\tempurl}


\bibitem[Jeunen(2019)]%
        {jeunen2019revisiting}
\bibfield{author}{\bibinfo{person}{Olivier Jeunen}.} \bibinfo{year}{2019}\natexlab{}.
\newblock \showarticletitle{Revisiting offline evaluation for implicit-feedback recommender systems}. In \bibinfo{booktitle}{\emph{Proceedings of the 13th ACM Conference on Recommender Systems}}. \bibinfo{pages}{596--600}.
\newblock


\bibitem[Jung et~al\mbox{.}(2020)]%
        {jung2020signed}
\bibfield{author}{\bibinfo{person}{Jinhong Jung}, \bibinfo{person}{Jaemin Yoo}, {and} \bibinfo{person}{U Kang}.} \bibinfo{year}{2020}\natexlab{}.
\newblock \showarticletitle{Signed graph diffusion network}.
\newblock \bibinfo{journal}{\emph{arXiv preprint arXiv:2012.14191}} (\bibinfo{year}{2020}).
\newblock


\bibitem[Kang and McAuley(2018)]%
        {kang2018self}
\bibfield{author}{\bibinfo{person}{Wang-Cheng Kang} {and} \bibinfo{person}{Julian McAuley}.} \bibinfo{year}{2018}\natexlab{}.
\newblock \showarticletitle{Self-attentive sequential recommendation}. In \bibinfo{booktitle}{\emph{2018 IEEE international conference on data mining (ICDM)}}. IEEE, \bibinfo{pages}{197--206}.
\newblock


\bibitem[Kim et~al\mbox{.}(2018)]%
        {kim2018side}
\bibfield{author}{\bibinfo{person}{Junghwan Kim}, \bibinfo{person}{Haekyu Park}, \bibinfo{person}{Ji-Eun Lee}, {and} \bibinfo{person}{U Kang}.} \bibinfo{year}{2018}\natexlab{}.
\newblock \showarticletitle{Side: representation learning in signed directed networks}. In \bibinfo{booktitle}{\emph{Proceedings of the 2018 world wide web conference}}. \bibinfo{pages}{509--518}.
\newblock


\bibitem[Kipf and Welling(2016)]%
        {kipf2016semi}
\bibfield{author}{\bibinfo{person}{Thomas~N Kipf} {and} \bibinfo{person}{Max Welling}.} \bibinfo{year}{2016}\natexlab{}.
\newblock \showarticletitle{Semi-supervised classification with graph convolutional networks}.
\newblock \bibinfo{journal}{\emph{arXiv preprint arXiv:1609.02907}} (\bibinfo{year}{2016}).
\newblock


\bibitem[Li et~al\mbox{.}(2020)]%
        {li2020learning}
\bibfield{author}{\bibinfo{person}{Yu Li}, \bibinfo{person}{Yuan Tian}, \bibinfo{person}{Jiawei Zhang}, {and} \bibinfo{person}{Yi Chang}.} \bibinfo{year}{2020}\natexlab{}.
\newblock \showarticletitle{Learning signed network embedding via graph attention}. In \bibinfo{booktitle}{\emph{Proceedings of the AAAI conference on artificial intelligence}}, Vol.~\bibinfo{volume}{34}. \bibinfo{pages}{4772--4779}.
\newblock


\bibitem[Luo et~al\mbox{.}(2024)]%
        {LUO}
\bibfield{author}{\bibinfo{person}{Huishi Luo}, \bibinfo{person}{Fuzhen Zhuang}, \bibinfo{person}{Ruobing Xie}, \bibinfo{person}{Hengshu Zhu}, \bibinfo{person}{Deqing Wang}, \bibinfo{person}{Zhulin An}, {and} \bibinfo{person}{Yongjun Xu}.} \bibinfo{year}{2024}\natexlab{}.
\newblock \showarticletitle{A survey on causal inference for recommendation}.
\newblock \bibinfo{journal}{\emph{The Innovation}} \bibinfo{volume}{5}, \bibinfo{number}{2} (\bibinfo{year}{2024}), \bibinfo{pages}{100590}.
\newblock
\showISSN{2666-6758}


\bibitem[Seo et~al\mbox{.}(2022)]%
        {seo2022siren}
\bibfield{author}{\bibinfo{person}{Changwon Seo}, \bibinfo{person}{Kyeong-Joong Jeong}, \bibinfo{person}{Sungsu Lim}, {and} \bibinfo{person}{Won-Yong Shin}.} \bibinfo{year}{2022}\natexlab{}.
\newblock \showarticletitle{SiReN: Sign-aware recommendation using graph neural networks}.
\newblock \bibinfo{journal}{\emph{IEEE Transactions on Neural Networks and Learning Systems}} (\bibinfo{year}{2022}).
\newblock


\bibitem[Shuman et~al\mbox{.}(2013)]%
        {shuman2013emerging}
\bibfield{author}{\bibinfo{person}{David~I Shuman}, \bibinfo{person}{Sunil~K Narang}, \bibinfo{person}{Pascal Frossard}, \bibinfo{person}{Antonio Ortega}, {and} \bibinfo{person}{Pierre Vandergheynst}.} \bibinfo{year}{2013}\natexlab{}.
\newblock \showarticletitle{The emerging field of signal processing on graphs: Extending high-dimensional data analysis to networks and other irregular domains}.
\newblock \bibinfo{journal}{\emph{IEEE signal processing magazine}} \bibinfo{volume}{30}, \bibinfo{number}{3} (\bibinfo{year}{2013}), \bibinfo{pages}{83--98}.
\newblock


\bibitem[Veli{\v{c}}kovi{\'c} et~al\mbox{.}(2017)]%
        {velivckovic2017graph}
\bibfield{author}{\bibinfo{person}{Petar Veli{\v{c}}kovi{\'c}}, \bibinfo{person}{Guillem Cucurull}, \bibinfo{person}{Arantxa Casanova}, \bibinfo{person}{Adriana Romero}, \bibinfo{person}{Pietro Lio}, {and} \bibinfo{person}{Yoshua Bengio}.} \bibinfo{year}{2017}\natexlab{}.
\newblock \showarticletitle{Graph attention networks}.
\newblock \bibinfo{journal}{\emph{arXiv preprint arXiv:1710.10903}} (\bibinfo{year}{2017}).
\newblock


\bibitem[Wang et~al\mbox{.}(2019)]%
        {wang2019neural}
\bibfield{author}{\bibinfo{person}{Xiang Wang}, \bibinfo{person}{Xiangnan He}, \bibinfo{person}{Meng Wang}, \bibinfo{person}{Fuli Feng}, {and} \bibinfo{person}{Tat-Seng Chua}.} \bibinfo{year}{2019}\natexlab{}.
\newblock \showarticletitle{Neural graph collaborative filtering}. In \bibinfo{booktitle}{\emph{Proceedings of the 42nd international ACM SIGIR conference on Research and development in Information Retrieval}}. \bibinfo{pages}{165--174}.
\newblock


\bibitem[Wu et~al\mbox{.}(2019)]%
        {wu2019simplifying}
\bibfield{author}{\bibinfo{person}{Felix Wu}, \bibinfo{person}{Amauri Souza}, \bibinfo{person}{Tianyi Zhang}, \bibinfo{person}{Christopher Fifty}, \bibinfo{person}{Tao Yu}, {and} \bibinfo{person}{Kilian Weinberger}.} \bibinfo{year}{2019}\natexlab{}.
\newblock \showarticletitle{Simplifying graph convolutional networks}. In \bibinfo{booktitle}{\emph{International conference on machine learning}}. PMLR, \bibinfo{pages}{6861--6871}.
\newblock


\bibitem[Wu et~al\mbox{.}(2022)]%
        {wu2022graph}
\bibfield{author}{\bibinfo{person}{Shiwen Wu}, \bibinfo{person}{Fei Sun}, \bibinfo{person}{Wentao Zhang}, \bibinfo{person}{Xu Xie}, {and} \bibinfo{person}{Bin Cui}.} \bibinfo{year}{2022}\natexlab{}.
\newblock \showarticletitle{Graph neural networks in recommender systems: a survey}.
\newblock \bibinfo{journal}{\emph{Comput. Surveys}} \bibinfo{volume}{55}, \bibinfo{number}{5} (\bibinfo{year}{2022}), \bibinfo{pages}{1--37}.
\newblock


\bibitem[Xu et~al\mbox{.}(2018)]%
        {xu2018powerful}
\bibfield{author}{\bibinfo{person}{Keyulu Xu}, \bibinfo{person}{Weihua Hu}, \bibinfo{person}{Jure Leskovec}, {and} \bibinfo{person}{Stefanie Jegelka}.} \bibinfo{year}{2018}\natexlab{}.
\newblock \showarticletitle{How powerful are graph neural networks?}
\newblock \bibinfo{journal}{\emph{arXiv preprint arXiv:1810.00826}} (\bibinfo{year}{2018}).
\newblock


\bibitem[Xu et~al\mbox{.}(2023)]%
        {xu2023artificial}
\bibfield{author}{\bibinfo{person}{Yongjun Xu}, \bibinfo{person}{Fei Wang}, \bibinfo{person}{Zhulin An}, \bibinfo{person}{Qi Wang}, {and} \bibinfo{person}{Zhao Zhang}.} \bibinfo{year}{2023}\natexlab{}.
\newblock \showarticletitle{Artificial intelligence for science—bridging data to wisdom}.
\newblock \bibinfo{journal}{\emph{The Innovation}} \bibinfo{volume}{4}, \bibinfo{number}{6} (\bibinfo{year}{2023}).
\newblock


\bibitem[Ying et~al\mbox{.}(2018)]%
        {ying2018graph}
\bibfield{author}{\bibinfo{person}{Rex Ying}, \bibinfo{person}{Ruining He}, \bibinfo{person}{Kaifeng Chen}, \bibinfo{person}{Pong Eksombatchai}, \bibinfo{person}{William~L Hamilton}, {and} \bibinfo{person}{Jure Leskovec}.} \bibinfo{year}{2018}\natexlab{}.
\newblock \showarticletitle{Graph convolutional neural networks for web-scale recommender systems}. In \bibinfo{booktitle}{\emph{Proceedings of the 24th ACM SIGKDD international conference on knowledge discovery \& data mining}}. \bibinfo{pages}{974--983}.
\newblock


\bibitem[Yuan et~al\mbox{.}(2017)]%
        {yuan2017sne}
\bibfield{author}{\bibinfo{person}{Shuhan Yuan}, \bibinfo{person}{Xintao Wu}, {and} \bibinfo{person}{Yang Xiang}.} \bibinfo{year}{2017}\natexlab{}.
\newblock \showarticletitle{SNE: signed network embedding}. In \bibinfo{booktitle}{\emph{Advances in Knowledge Discovery and Data Mining: 21st Pacific-Asia Conference, PAKDD 2017, Jeju, South Korea, May 23-26, 2017, Proceedings, Part II 21}}. Springer, \bibinfo{pages}{183--195}.
\newblock


\bibitem[Zhang et~al\mbox{.}(2018)]%
        {zhang2018coupledcf}
\bibfield{author}{\bibinfo{person}{Quangui Zhang}, \bibinfo{person}{Longbing Cao}, \bibinfo{person}{Chengzhang Zhu}, \bibinfo{person}{Zhiqiang Li}, {and} \bibinfo{person}{Jinguang Sun}.} \bibinfo{year}{2018}\natexlab{}.
\newblock \showarticletitle{Coupledcf: Learning explicit and implicit user-item couplings in recommendation for deep collaborative filtering}. In \bibinfo{booktitle}{\emph{IJCAI International Joint Conference on Artificial Intelligence}}.
\newblock


\bibitem[Zhang et~al\mbox{.}(2023)]%
        {zhang2023contrastive}
\bibfield{author}{\bibinfo{person}{Zeyu Zhang}, \bibinfo{person}{Jiamou Liu}, \bibinfo{person}{Kaiqi Zhao}, \bibinfo{person}{Song Yang}, \bibinfo{person}{Xianda Zheng}, {and} \bibinfo{person}{Yifei Wang}.} \bibinfo{year}{2023}\natexlab{}.
\newblock \showarticletitle{Contrastive learning for signed bipartite graphs}. In \bibinfo{booktitle}{\emph{Proceedings of the 46th International ACM SIGIR Conference on Research and Development in Information Retrieval}}. \bibinfo{pages}{1629--1638}.
\newblock


\bibitem[Zhou et~al\mbox{.}(2020)]%
        {zhou2020s3}
\bibfield{author}{\bibinfo{person}{Kun Zhou}, \bibinfo{person}{Hui Wang}, \bibinfo{person}{Wayne~Xin Zhao}, \bibinfo{person}{Yutao Zhu}, \bibinfo{person}{Sirui Wang}, \bibinfo{person}{Fuzheng Zhang}, \bibinfo{person}{Zhongyuan Wang}, {and} \bibinfo{person}{Ji-Rong Wen}.} \bibinfo{year}{2020}\natexlab{}.
\newblock \showarticletitle{S3-rec: Self-supervised learning for sequential recommendation with mutual information maximization}. In \bibinfo{booktitle}{\emph{CIKM}}.
\newblock


\end{thebibliography}
\appendix
\let\cleardoublepage\relax
\section{Experimental Setting}
\subsection{Dataset}
\label{app:dataset}
In this paper, to evaluate the performance of the proposed DFGNN, we conduct comprehensive experiments on classical ML1M, different categories of Amazon Review datasets, and Yelp dataset. Amazon Review datasets are collected from the Amazon shopping website. Each instance consists of a user's rating for an item, ranging from 1 to 5.  we  select three categories dataset, they are \textit{ArtsCrafts and Sewing},\textit{Grocery and Gourmet Food}, and \textit{ Kindle Store}. For all datasets, we regard the ratings that are below 3 as negative feedback, and the ratings that are higher than 3 as positive feedback. Following previous works\cite{kang2018self,zhou2020s3},
we discard the users and items with less than five interactions. For each dataset we randomly select 70\% instances as training data, 10\% instances as validation data, and 20\% instances as testing data. The detailed statistical information of those datasets is shown in Tab.~\ref{tab:dataset} 
\begin{table}[!htbp]
\small
\caption{Statistics of all datasets.}
\label{tab:dataset}
\center
\begin{tabular}{l|rrrrrr}
\toprule
Dataset &\# user&\# item& \# instance& \# negative rate\\
\midrule

\multirow{1}{*}{\ Arts}
~&56123 & 22847 & 413554 & 5.85\%\\
% \multirow{1}{*}{\ Office}
% ~&101454 & 27918 & 692857 & 7.44\%\\
\multirow{1}{*}{\ Kindle}
~&139729 & 98821 & 2020496 & 5.68\%\\
\multirow{1}{*}{\ GFood}
~&127399 & 41317 & 990360 & 8.74\%\\
\multirow{1}{*}{\ Yelp}
~&207453 & 92422 & 3139963 & 19.94\%\\
\multirow{1}{*}{\ ML1M}
~&6040 & 3668 & 739012 & 22.16\%\\
\bottomrule
% \multirow{1}{*}{\ ML-1M}
% ~&6040&3707&1,000,209&269.89\\
% \bottomrule
\end{tabular}
\end{table}

\subsubsection{Implement Details}
\label{app:Implement details}.
For all the models the embedding size of items and users is set to 64 and the batch size is set to 512. 
The GNN layer K is set to 2 in our method.
For all methods. We optimize all models with the Adam Optimizer and carefully search for the hyper-parameters of all baselines. To avoid overfitting,  we adopt the early stop strategy with a patience of 20 epochs. For the recommendation ranking task, we adopt the random negative sampling strategy with a 1:1 sampling rate.   We conduct a grid search for hyper-parameters. We search for models’ learning rates among $\{1e^-2,3e^{-3},1e^{-3}, 3e^{-4}, 1e^{-4}, 3e^{-5}\}$. 

\section{Baselines}
 % To validate our IDP, we conduct a comprehensive comparison. Specifically, we compare our method with the following baselines:(1)\textbf{ GRU4Rec}\cite{hidasi2016session} (2)\textbf{ SASRec} \cite{kang2018self}, (3)\textbf{ BERT4Rec} \cite{sun2019BERT4Rec}, (4)\textbf{ FDSA} \cite{zhang2019feature}, (5)\textbf{Recformer}\cite{li2023text}, and (6)\textbf{UniSRec}$_{ID+t}$ \cite{hou2022towards}. Please refer to the \textbf{Appendix.\ref{Baselines}} for a detailed description of baselines.

\label{Baselines}
  To validate our IDP, we conduct a comprehensive comparison. Specifically, we compare our method with the following baselines: (1)\textbf{ GCN}\cite{kipf2016semi}, (2)\textbf{ GAT} \cite{velivckovic2017graph}, (3) \textbf{SGCN}\cite{derr2018signed} (4)\textbf{SBGNN} \cite{huang2021signed}, (5)\textbf{ SBGCL} \cite{zhang2023contrastive}. 
    % \textbf{ GRU4Rec} \cite{hidasi2016session}, which adopts GRU as  user sequence encoder for sequential recommenadtion. 
    %  \textbf{ SASRec} \cite{kang2018self}, which is a classical sequential recommendation model that introduces self-attention in behavior modeling. 
    %  \textbf{ BERT4Rec} \cite{sun2019BERT4Rec}, which adopts masked item prediction as an optimizing object.
    %  \textbf{ FDSA} \cite{zhang2019feature}, which adopts self-attentive networks to model item and feature transition. In this paper, we adopt item descriptions as item features. 
    %  \textbf{UniSRec}$_{ID+t}$ \cite{hou2022towards}, UniSRec is a SOTA text-based pre-train model. In the pre-traning stage, UniSRec represents items with text embedding encoded by a pre-trained language model and pre-train recommendation model based on text embedding. In the downstream domain, UniSRec fine-tunes pre-trained model by the designed MoE-enhanced adopter. UniSRec$_{ID+t}$ is the final version of UniSRec, which add ID embedding in downstream domains.  

 \begin{itemize}[leftmargin=*]
       \item { GCN} \cite{kipf2016semi},which is a classical graph convolution network. GCN adopts the first-order Chebyshev polynomial to construct graph filters.
     \item{ GAT} \cite{velivckovic2017graph}, which is a classical attention-based GNN. It adopts the attention mechanism to aggregation neighbor nodes.

     \item{ SGCN} \cite{derr2018signed}, which is a classical signed graph convolution network. It carefully designs a message-passing mechanism based on balanced theory.
     \item{ SBGNN} \cite{huang2021signed}, which is designed for bipartite signed graph. It extends balance theory to bipartite graphs by butterfly structure.
     \item {SBGCL} \cite{zhang2023contrastive},which is a contrastive based signed graph neural network. It improved the graph contrastive learning to adapt to signed graphs. 

    We implement our method by PyG, which is a popular graph learning framework. For baseline GCN, GAT, and SGCN we directly use the implementation of PyG. For baseline SBGNN and SBGCL, we directly use the author's released codes.
 
 \end{itemize}
  
\end{document}